%% file: IEEE-conference-template-062824.tex
\documentclass[10pt,conference]{IEEEtran}

\IEEEoverridecommandlockouts
\usepackage{booktabs} 
\usepackage{multirow} 
\usepackage{graphicx} 
\usepackage{cite}
\usepackage{amsmath,amssymb,amsfonts}
\usepackage{algorithmic}
\usepackage{graphicx}
\usepackage{textcomp}
\usepackage{xcolor}
\usepackage{microtype}
\usepackage{booktabs} 
\usepackage{float} 
\usepackage{subfigure} 
\usepackage{tcolorbox}
\usepackage{balance}
\usepackage{enumitem}
\usepackage{graphicx} 
\usepackage{subcaption} 
\usepackage{hyperref}
\usepackage{pifont}  
\usepackage{scalerel}
\usepackage{xspace}
\usepackage{subcaption}
\hypersetup{
  colorlinks,
  linkcolor=black, 
  urlcolor=blue,
  citecolor=black,
  bookmarksnumbered,
  unicode
}
\newtcolorbox{findingbox}{
  colback=lightgray!20,
  colframe=gray!50,
  boxrule=0.5pt,
  left=10pt,
  right=10pt,
  top=10pt,
  bottom=10pt,
  arc=0pt,
  boxsep=0pt
}
\definecolor{dimGreen}{rgb}{0.3, 0.6, 0.3} 

\def\BibTeX{{\rm B\kern-.05em{\sc i\kern-.025em b}\kern-.08em
    T\kern-.1667em\lower.7ex\hbox{E}\kern-.125emX}}

\begin{document}

\title{Empirical Evaluation of Large Language Models in Automated Program Repair}
\author{%
    \IEEEauthorblockN{Jiajun Sun\textsuperscript{1}, Fengjie Li\textsuperscript{1}, Xinzhu Qi\textsuperscript{2}, Hongyu Zhang\textsuperscript{3}} and Jiajun Jiang\textsuperscript{1}\thanks{$^*$Jiajun Jiang is the corresponding author for this work.} 
    \IEEEauthorblockA{%
    \textsuperscript{1}College of Intelligence and Computing, Tianjin University, China\\
    \textsuperscript{2}School of Information and Software Engineering, University of Electronic Science and Technology of China, China\\
    \textsuperscript{3}School of Big Data and Software Engineering, Chongqing University, China \\
    \{sjjtianjin, fengjie\}@tju.edu.cn, 2022090905024@std.uestc.edu.cn, hyzhang@cqu.edu.cn, jiangjiajun@tju.edu.cn
    }
}
\pagestyle{plain}

\maketitle
\input{macros}
\begin{abstract}

The increasing prevalence of software bugs has positioned automated program repair (APR) as a critical area of research. The rapid evolution of large language models (LLMs) offers new opportunities for advancing APR techniques. While prior studies have explored LLMs in repair tasks, most have relied on earlier-generation, smaller models and focused predominantly on Java-based benchmarks. The repair capabilities of modern, large-scale LLMs across diverse programming languages and scenarios remain insufficiently examined. To address this gap, we conducted a comprehensive empirical study evaluating four representative open-source LLMs: CodeLlama, LLaMA, StarCoder, and DeepSeek-Coder, spanning a range of model sizes (7B to 33B parameters), architectural designs, and application scopes (general purpose vs. code-specialized). Our evaluation spans (1) two distinct bug scenarios (enterprise-grades and algorithmic assignment bugs), (2) three programming languages (Java, C/C++, and Python), and (3) four prompt engineering strategies, resulting in the generation and analysis of over 600,000 patches across six benchmark datasets. Our findings reveal several key insights. First, although larger models tend to yield better repair performance, the fine-tuned CodeLlama consistently outperforms the general-purpose LLaMA across multiple benchmarks, despite having fewer parameters. Second, successful repair does not scale linearly with model size; smaller models can generate unique correct patches not found in larger models. Third, correct patches are frequently produced in the early stages of patch generation. Finally, prompt design significantly influences repair effectiveness.These findings underscore the value of model specialization over sheer scale, advocate for cost-aware verification mechanisms, and highlight the importance of prompt engineering in LLM-driven APR. Our study provides actionable guidance for building practical, efficient, and effective APR systems using LLMs.

\end{abstract}


\input{1Introduction}

\input{7RelatedWork}

\input{3StudyDesign}

\input{5ResultAnalysis}

\input{6Discussion}

\input{8Conclusion}

\balance
\bibliographystyle{IEEEtran}
\bibliography{IEEEabrv,reference}
\end{document}

%% file: macros.tex
\newcommand{\jiajun}[1]{\textcolor{blue}{[Jiajun: #1]}}

\definecolor{hycolor}{rgb}{0.7,0.7,0.3} 
\newcommand{\nbc}[3]{
	{\colorbox{#3}{\bfseries\sffamily\scriptsize\textcolor{white}{#1}}}
	{\textcolor{#3}{\sf\small$\blacktriangleright$\emph{#2}$\blacktriangleleft$}}
}
\newcommand{\hy}[1]{\nbc{HY}{#1}{hycolor}}

\newcommand{\fj}[1]{\textcolor{orange}{[Fengjie: #1]}}

\newcommand{\sjj}[1]{\textcolor{purple}{[sjj: #1]}}

\newcommand{\codeIn}[1]{{\ttfamily #1}}

\newcommand{\dfjvone}{Defects4J v1.2}
\newcommand{\dfjvtwo}{Defects4J v2.0}
\newcommand{\dfj}{Defects4J}
\newcommand{\bugscpp}{BugsCpp}

\newcommand{\introclass}{IntroClass}
\newcommand{\introclassc}{IntroClass-C}
\newcommand{\introclassjava}{IntroClass-Java}

\newcommand{\condefects}{ConDefects}
\newcommand{\condefectsjava}{ConDefects-Java}
\newcommand{\condefectspython}{ConDefects-Py}

\newcommand{\llama}{LLaMA}
\newcommand{\codellama}{CodeLlama}
\newcommand{\starcoder}{StarCoder}
\newcommand{\deepseekcoder}{DeepSeek-Coder}

\newcommand{\openai}{\scalerel*{\includegraphics{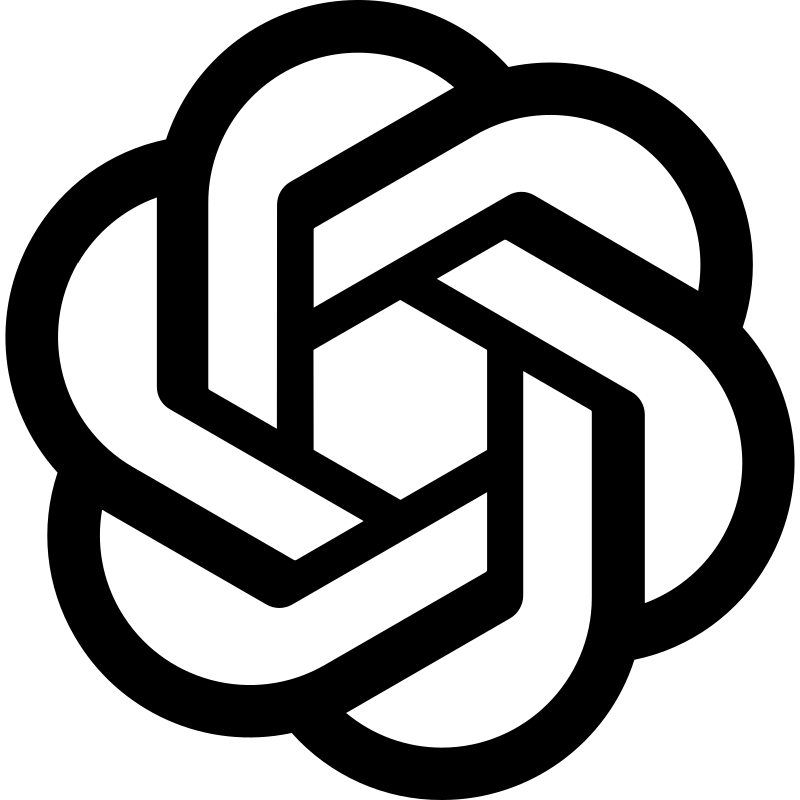}}{\textrm{C}}\xspace}
\newcommand{\bigcode}{\scalerel*{\includegraphics{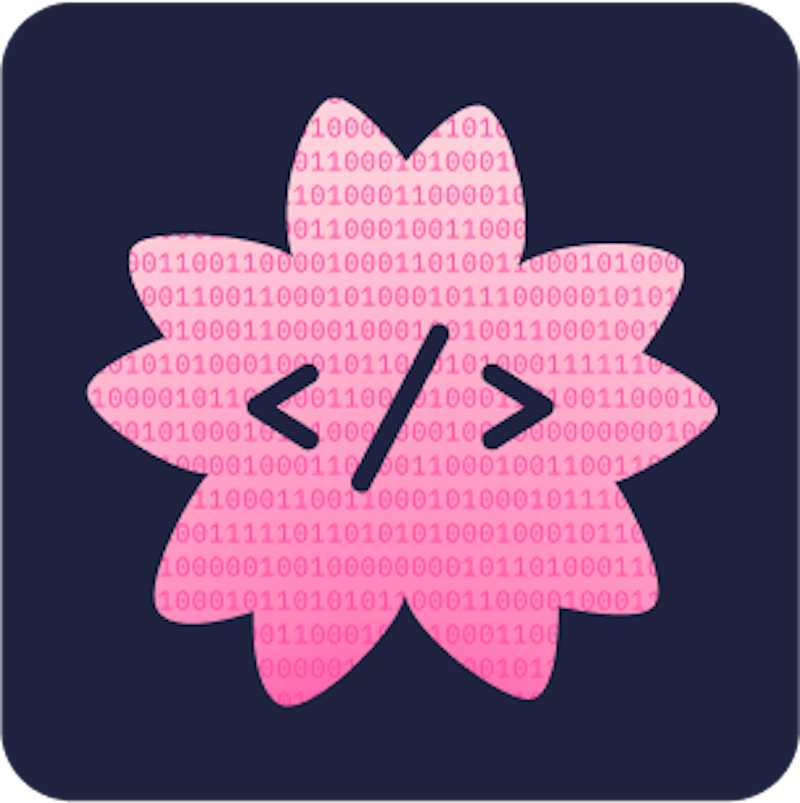}}{\textrm{C}}\xspace}
\newcommand{\meta}{\scalerel*{\includegraphics{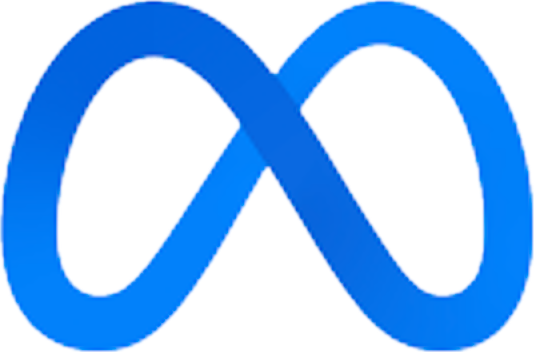}}{\textrm{C}}\xspace}
\newcommand{\deepseek}{\scalerel*{\includegraphics{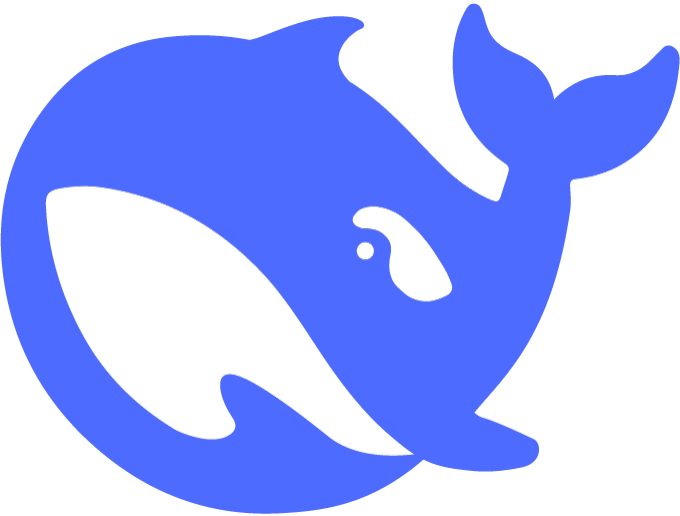}}{\textrm{C}}\xspace}

%% file: 1Introduction.tex
\section{Introduction}
\label{sec:Introduction}

In recent years, as software systems have grown in scale and complexity, software bugs have become more prevalent, leading to substantial maintenance costs\cite{o2017debugging}. Automated Program Repair (APR), which aims to generate correct patches automatically to reduce developer effort, has garnered considerable attention from both academia and industry. Traditional APR approaches~\cite{le2011genprog,xuan2016nopol,le2017s3,long2015staged,mechtaev2016angelix,martinez2016astor,Le2016HistoryDP,long2016automatic,xin2017leveraging,xiong2017precise,long2017automatic,hua2018towards, wen2018context,jiang2018shaping,liu2019tbar,liu2019avatar,ghanbari2019practical,jiang2019inferring}, such as heuristic-based methods~\cite{le2011genprog, Le2016HistoryDP,long2016automatic,xin2017leveraging,xiong2017precise,wen2018context,jiang2018shaping}, constraint-driven methods~\cite{xuan2016nopol,le2017s3,long2015staged,mechtaev2016angelix} and template-based methods~\cite{martinez2016astor,long2017automatic,hua2018towards,ghanbari2019practical,jiang2019inferring,liu2019avatar}, have shown promising results but remain constrained by their reliance on domain-specific rules, poor scalability, and difficulty in generalizing to diverse types of bugs~\cite{xia2023plastic}. In comparison, deep learning-based methods~\cite{chen2019sequencer,lutellier2020coconut,li2020dlfix, zhu2021syntax, jiang2021cure, ye2022neural, zhu2023tare} trained on large code-patch corpora have shown advanced performance. However, despite their progress in addressing a broader range of bugs, these methods remain constrained by the scope of their training data, which typically cover only a limited spectrum of bug types, restricting their ability to generalize to previously unseen or rare bug types~\cite{fu2022vulrepair,Xia2022LessTM,feng2024prompting}.

More recently, large language models (LLMs) have demonstrated impressive capabilities across various software engineering tasks~\cite{guo2022unixcoder,nijkamp2022codegen,wang2023codet5+,ahmed2022few}, showing their profound code comprehension and generative capabilities. Several recent studies~\cite{prenner2022can, jiang2023impact, xia2023automated, fan2023automated} have also explored the application of LLMs to APR tasks, showing promising results. Following these studies, numerous LLM-based APR techniques~\cite{Xia2022LessTM, zhang2023gamma, xia2023plastic, yin2024thinkrepair, xia2023keep} have been proposed, further exploring the potential of LLMs in APR tasks.

However, existing studies~\cite{huang2023empirical,fan2023automated,jiang2023impact, xia2023automated, xiang2024far} either predominantly investigate smaller-scale and earlier-generation models~\cite{huang2023empirical,fan2023automated,jiang2023impact, xia2023automated} (e.g., CodeBERT~\cite{feng2020codebert}, CodeT5~\cite{wang2021codet5}), or primarily evaluate on Java benchmarks~\cite{xiang2024far}, particularly the Defects4J~\cite{just2014defects4j} dataset. The exploration of newer, larger-scale LLMs across multiple programming languages and datasets remains limited, thereby leaving the following several key aspects insufficiently investigated: \textbf{1) Multilingual Generalizability}: Although some existing studies~\cite{huang2023empirical, xia2023automated} have evaluated LLMs on C/C++ or Python datasets, the scale of these benchmarks is typically limited (e.g. only 91 bugs from ManyBugs~\cite{le2015manybugs} and 40 bugs from QuixBugs~\cite{lin2017quixbugs}). Moreover, these studies often lack a detailed analysis of LLM performance across different programming languages, leaving the multilingual repair capabilities of modern LLMs largely underexplored. \textbf{2) Cost-Effectiveness Tradeoffs}: As the parameter scales of large models continue to expand, the associated computational costs have become increasingly significant~\cite{tu2024overview,sharir2020cost}. However, existing works rarely examine the tradeoffs between repair effectiveness and resource consumption, which is crucial for determining practical deployment settings in real-world scenarios. \textbf{3) Unexplored Influential Factors}: these existing works primarily focus on the performance of LLMs via fine-tuning or prompt-engineering strategies, yet fall short of offering systematic investigations of other influencing factors, including intrinsic properties of LLMs, the characteristics of datasets and the features of generated patches.

To address these gaps, this paper comprehensively investigates the repair capabilities of popular LLMs across multiple scenarios, programming languages, and prompt engineering strategies, aiming to provide actionable insights for developing more practical APR tools and to further bridge the gaps between LLMs and APR. Specifically, we evaluate four representative open-source LLMs, \codellama{}~\cite{roziere2023code}, \llama{}~\cite{touvron2023llama}, \starcoder{}~\cite{li2023starcoder} and \deepseekcoder{}~\cite{guo2024deepseekcoderlargelanguagemodel}, covering a range of parameter sizes from 7B to 33B. Among these LLMs, \llama{} serves as a general-purpose LLM, while the other three are code-specialized LLMs trained on programming-related tasks. We evaluate these four LLMs across two critical bug repair scenarios (enterprise-grade project bugs and algorithmic competition bugs) and three widely used programming languages (Java, C/C++, Python). Our evaluation leverages six benchmarks: \dfj{}~\cite{just2014defects4j}, \bugscpp{}~\cite{10298287}, \introclass{}~\cite{le2015manybugs}, \introclassjava{}~\cite{durieux2016introclassjava}, \condefectsjava{}~\cite{wu2023ConDefectsnewdatasetaddress}, and \condefectspython{}~\cite{wu2023ConDefectsnewdatasetaddress}. In addition, we design four distinct prompting strategies to systematically explore how prompt content influences repair effectiveness. Through \textbf{rigorously analyzing over 600,000 generated patches}, our study provides several important insights into the performance and behavior of LLMs in APR:

\textbf{Repair Effectiveness}. Our results show that LLM demonstrates stronger performance on fixing algorithmic competition bugs compared to enterprise-grade project bugs. For instance, the best-performing model DeepSeek achieves an average repair rate of 45.45\% on the algorithm competition benchmarks IntroClass, while its average repair rate on enterprise-grade projects bugs benchmarks BugsCpp is significantly lower at 5.66\%. Moreover, we find that fine-tuning on code-related tasks significantly enhances LLMs’ repair capabilities: the fine-tuned CodeLlama-7B model consistently outperforms the larger LlaMa-2-13B across all evaluated datasets despite having fewer parameters.

\textbf{Computational Costs}. For code-specialized models, increasing parameter size yields diminishing returns in repair effectiveness; that is, doubling the number of parameters does not lead to proportionally greater gains in the number of correct patches. For example, despite having 4.71 times the number of parameters as CodeLlama, DeepSeek only repaired 4 fewer bugs than CodeLlama on the Defects4J 1.2 dataset. Interestingly, smaller-scale models exhibit complementary repair capabilities that are not entirely subsumed by their larger counterparts. For example, CodeLlama can fix 9 bugs that cannot be fixed by any other model. Furthermore, our analysis shows that correct patches are predominantly generated in early iterations of generation. For example, our analysis of StarCoder's performance on the IntroClassJava benchmark revealed that 95.77\% of correct patches were generated within the first 30 generations. In a similar manner, 100\% of correct patches generated by LLaMA on both the IntroClass and IntroClassJava benchmarks were produced within the first 7 generations. This insight suggests practical strategies for optimizing computational resources and improving repair efficiency, such as prioritizing early-stage outputs or leveraging a combination of models of varying scales.

\textbf{Prompt Engineering}. The evaluated LLMs exhibit high sensitivity to variations in prompt format. Incorporating repair examples into prompts significantly improves repair success rates, particularly for weaker models. For instance, on the ConDefectsPy dataset, when a repair example is provided in the prompt, CodeLlama's repair count increases by \textbf{206.7\%} compared to when no repair example is included in the prompt. Additionally, integrating bug analysis generated by LLM into prompts further enhances repair performance and contributes additional unique fixes. CodeLlama successfully repairs 51 bugs that it failed to repair when error analysis was not given. Flawed or inaccurate diagnostic information can adversely affect the accuracy of stronger models. However, the LLM's error analysis incorrectly identified the issue as the absence of a check to determine whether A and B are divisible by 10. This erroneous analysis led to a situation where DeepSeek-Coder could have potentially completed the repair, but instead, DeepSeek only modified the code based on the incorrect reason, thus failing to achieve the correct fix. It highlights the importance of carefully validating auxiliary content. These findings highlight the critical role of systematic prompt optimization in LLM-based APR workflows.

To sum up, this paper makes the following contributions:
\begin{enumerate}[label={\textbullet}]
    \item We conduct the first systematic evaluation of four representative open-source LLMs (LLaMA, CodeLlama, StarCoder and DeepSeek-Coder) across six diverse benchmarks covering two representative repair scenarios (enterprise-grade project bugs and algorithmic competition bugs) and three popular programming languages (Java, C/C++, Python).
    \item We conduct an in-depth analysis over \textbf{600,000 generated patches}, and we provide empirical insights into how model architecture, dataset characteristics, and prompt engineering strategies influence repair effectiveness.

    \item Our findings reveal that smaller-scale LLMs can produce unique correct patches that are not recoverable by larger models, and that most correct patches emerge in the early stages of generation. These findings inform practical guidelines for balancing computational costs and repair performance in real-world deployments.
    \item We employ and evaluate four distinct prompting strategies to assess their impact on the repair capabilities of LLMs, offering practical insights and actionable guidelines for optimizing prompt usage in future APR workflow designs.
    
    \item To foster reproducibility and future development, we release all generated patches, evaluation scripts, prompt templates, and dataset configurations at our homepage.

\end{enumerate}

%% file: 7RelatedWork.tex
\section{Background and Related Work}
\label{sec:back_and_related}

\subsection{Large Language Models}

Large Language Models (LLMs) contain millions to billions of parameters and are pretrained on large-scale corpora encompassing natural language, programming language, mathematical problems, and other diverse domains. Modern LLMs predominantly adopt the Transformer~\cite{10.5555/3295222.3295349} architecture, which utilizes self-attention mechanisms to capture contextual relationships between tokens. The Transformer typically follows an encoder-decoder structure: the encoder transforms input sequences into contextualized representations, while the decoder generates output sequences by attending to both the encoder’s outputs and previously generated tokens. According to their underlying architectures,  LLMs can be classified into three primary categories: encoder-only models, decoder-only models and encoder-decoder models. Among them, decoder-only LLMs have demonstrated superior capabilities in text comprehension and generation~\cite{kaplan2020scaling}, attracting significant attention from both academia and industry. To adapt LLMs for specific downstream tasks, many works apply fine-tuning~\cite{touvron2023llama,sun2019fine,ding2023parameter} or prompt-tuning~\cite{lester2021power,jia2022visual} techniques. For instance, \codellama{}~\cite{roziere2023code} is derived from \llama{}~\cite{touvron2023llama} through fine-tuning on code-related tasks, while RepairLlama~\cite{silva2023repairllama} further applies LoRA-based~\cite{hu2022lora} fine-tuning on \codellama{} for the APR task. In addition, many approaches~\cite{brown2020language,wei2022chain,zhang2022auto,wang2023towards} leverage few-shot learning~\cite{brown2020language} or chain-of-thought (COT)~\cite{wei2022chain,zhang2022auto,wang2023towards} prompting to inject task-specific knowledge into models via carefully designed prompts, guiding the model to better understand and resolve targeted problems.

\subsection{Automated Program Repair}
Automated Program Repair (APR) aims to automatically fix software bugs while reducing manual effort. Traditional APR approaches primarily fall into three categories: (1) Constraint-based methods~\cite{xuan2016nopol,le2017s3,long2015staged,mechtaev2016angelix} leverage symbolic execution and constraint solving to extract semantic information for improved patch generation. (2) Heuristic-based techniques~\cite{le2011genprog, Le2016HistoryDP,long2016automatic,xin2017leveraging,xiong2017precise,wen2018context,jiang2018shaping} employ genetic programming or stochastic search strategies to expand the search space and enhance correct patch identification. (3) Template-based approaches~\cite{martinez2016astor,long2017automatic,hua2018towards,ghanbari2019practical,jiang2019inferring,liu2019avatar}, currently demonstrating the highest effectiveness in research, utilize predefined repair patterns either manually designed or mined from historical fixes. However, their inherent limitation lies in the inability to repair bugs beyond the covered templates. These traditional approaches struggle with the large and diverse search space of bugs, resulting in a limited number of fixes. Deep learning has opened new possibilities: neural APR methods~\cite{chen2019sequencer,lutellier2020coconut,li2020dlfix, zhu2021syntax, jiang2021cure, ye2022neural, zhu2023tare}, trained on massive code-patch pairs can significantly improve repair effectiveness, yet their performance often depends on the quality and coverage of training data.

Recently, numerous LLM-based APR methods have been proposed. AlphaRepair~\cite{Xia2022LessTM} first leverages CodeBERT to replace masked tokens in the buggy code to generate patches. FitRepair~\cite{xia2023plastic} leverages the plastic surgery hypothesis to enhance repair performance by fine-tuning CodeT5 models. Similarly, Repilot~\cite{wei2023copiloting} combines a code completion engine with CodeT5 to regulate the syntactic correctness of generated patches and improve repair effectiveness. Additionally, GAMMA~\cite{zhang2023gamma} employs CodeBERT and UniXcoder to fill a set of predefined repair templates to generate patches. More recently, with the popularity of larger-scale models such as ChatGPT~\cite{NEURIPS2020_1457c0d6}, several new APR approaches have been proposed, including ChatRepair~\cite{xia2023keep}, ThinkRepair~\cite{yin2024thinkrepair}, and SRepair~\cite{xiang2024far}. These methods leverage LLMs with stronger code understanding and generation capabilities and reformulate the APR task as one of effectively supplying the LLM with relevant knowledge and feedback to facilitate the repair process. However, most of these approaches focus solely on the performance on the Defects4J dataset, lacking a more comprehensive evaluation of these LLMs across diverse benchmarks.

\subsection{Existing Studies on LLM-based APR}
\label{sec:study_on_LLM_APR}

Several studies have already explored the use of LLMs for APR. One similar work is Xia et al.~\cite{xia2023automated}, which was the first to apply LLMs to the APR task, evaluating different LLMs on Defects4J, ManyBugs and QuixBugs datasets. Although these datasets cover multiple programming languages (Java, Python and C), their overall scale and diversity are limited (e.g., QuixBugs contains only 40 simple python bugs, and only 91 C bugs from ManyBugs were used in this study). While their study demonstrated the promising potential of LLMs, the models employed were mostly earlier-generation LLMs with relatively small parameter sizes. Moreover, the study lacked a comprehensive analysis of the performance of LLMs across different programming languages and did not provide an in-depth analysis of LLM-generated patches. Although Fan et al.~\cite{fan2023automated} and Xiang et al.~\cite{xiang2024far} employed more recent LLMs in their studies, their evaluation datasets included only bugs written in Java. Besides some studies~\cite{jiang2023impact, huang2023empirical} have explored enhancing the performance of smaller-scale models (e.g. CodeBert and CodeT5) on APR tasks through fine-tuning techniques. Different from these studies, our work employs four modern open-source LLMs that are widely adopted in contemporary research. Second, while prior C/C++ analyses focused exclusively on vulnerability-specific datasets, our evaluation incorporates both enterprise-grade project bugs and algorithmic error cases across three mainstream languages: Java, C/C++, and Python. Furthermore, we conduct a comprehensive analysis spanning multiple dimensions: 1) dataset characteristics, 2) model scale effects, and 3) prompt design considerations, providing holistic insights into LLM capabilities for software repair. 

Unlike these studies, our work leverages six diverse datasets comprising \textbf{a total of 2,309 bugs from both enterprise-grade projects and algorithmic assignments}, spanning three popular programming languages (Java, Python, C/C++). We evaluate four modern open-source LLMs widely used in current research~\cite{li2025bybrid,silva2023repairllama,yin2024thinkrepair}. Furthermore, beyond simply comparing the repair performance of different LLMs, we conduct an in-depth analysis of the underlying factors that influence their effectiveness, including: dataset characteristics, parameter sizes, patch generation behaviors, and the impact of prompt design. Our findings offer valuable insights for effectively leveraging LLMs in APR.

%% file: 3StudyDesign.tex
\section{Study Design}
\subsection{Research Questions}
\label{sec:RQ}
To provide a comprehensive analysis of the capabilities of LLMs in APR, as well as the factors influencing repair effectiveness, we answer the following research questions:

\textbf{RQ1: How effective are LLMs in repairing bugs from enterprise-grade projects?} We employed four selected LLMs to repair bugs from the Defects4J and BugsCpp datasets, enabling a comprehensive assessment of the current capabilities of these LLMs in fixing enterprise-grade project bugs. By further analyzing the similarities and differences in their repair behaviors across these datasets, we aim to identify factors that influence their effectiveness, such as programming language, project characteristics, and model characteristics.

\textbf{RQ2: How effective are LLMs in repairing bugs in algorithmic assignments?} Similar to the previous question, we utilized the selected LLMs to repair bugs across four algorithmic assignment datasets, providing a comprehensive evaluation of the capabilities of current popular LLMs in addressing bugs in such tasks. Furthermore, we analyzed the ranking of correct patches to investigate how to balance repair effectiveness with the associated generation cost.

\textbf{RQ3: Impact of Bug Length and Patch Actions.} From the findings of the previous research questions, we observed that the repair performance of the studied LLMs varies significantly across different programming languages and types of datasets. In this RQ, we investigate potential factors contributing to this phenomenon, including the characteristics of bugs and patches in each dataset and across different programming languages.

\textbf{RQ4: Impact of Prompt Settings on Repair Performance.} In this RQ, we investigate the impact of different prompt settings on the repair effectiveness of studied LLMs. Specifically, we collect three commonly used prompt settings from existing APR literature and systematically study how each affects the repair performance of LLMs.

\subsection{Datasets}
\input{tables/dataset}
In our study, we selected two types of datasets:
\begin{itemize}
    \item  \textbf{Enterprise-grade project bugs:} Defects4J~\cite{just2014defects4j} and BugsCpp~\cite{10298287} contain 835 and 215 bugs from real-world projects, respectively. In this paper, we chose 255 and 228 single-function bugs from \dfjvone{} and v2.0, respectively, along with 106 single-function bugs from \bugscpp{}.
    
    \item  \textbf{Algorithmic assignment bugs:} The \introclass{}~\cite{le2015manybugs} dataset consists of 998 bugs found in student-written implementations of six small C programming assignments. The \introclassjava{}~\cite{durieux2016introclassjava} dataset is constructed by translating each \introclass{} program into Java, preserving the original code structure and semantics. To distinguish between programming languages, we refer to the \introclass{} dataset as \introclassc{} in this paper, and we consider 297 single-function bugs from each of these two datasets. The \condefects{}~\cite{wu2023ConDefectsnewdatasetaddress} dataset was collected from failed submissions to programming assignments on the AtCoder platform~\cite{atcoder}, containing 1,254 Java bugs and 1,625 Python bugs. To distinguish between programming languages, we refer to the two subsets as \condefectsjava{} and \condefectspython{} in this paper. To ensure semantic consistency between the two datasets, we selected the programming assignment that contains bugs in both Java and Python versions. Since multiple bugs may exist for a single language under an assignment, we randomly selected one faulty submission per programming language to reduce experimental overhead. In total, we retained 563 single-function bugs in each dataset.
\end{itemize}

\subsection{Large Language Models}
\input{tables/llm}

As shown in Table~\ref{tab:models}, we selected four open-source LLMs, including a general-purpose model, \llama{}-2-13B, and three models specifically designed for code-related tasks: \codellama{}-7B, which is derived from \llama{}-2-7B and further fine-tuned on code-related tasks, as well as \starcoder{}Base and \deepseekcoder{}-33B-instruct. The selected models aim to cover diverse parameter sizes and model providers and include both general-purpose and code-specialized LLMs that are widely used in existing research, ensuring a representative and comprehensive comparison.

\subsection{Prompt Template}

\begin{figure}[tbp] 
\centering 
\includegraphics[width=0.49\textwidth,trim=0cm 1cm 2.5cm 1cm,clip]{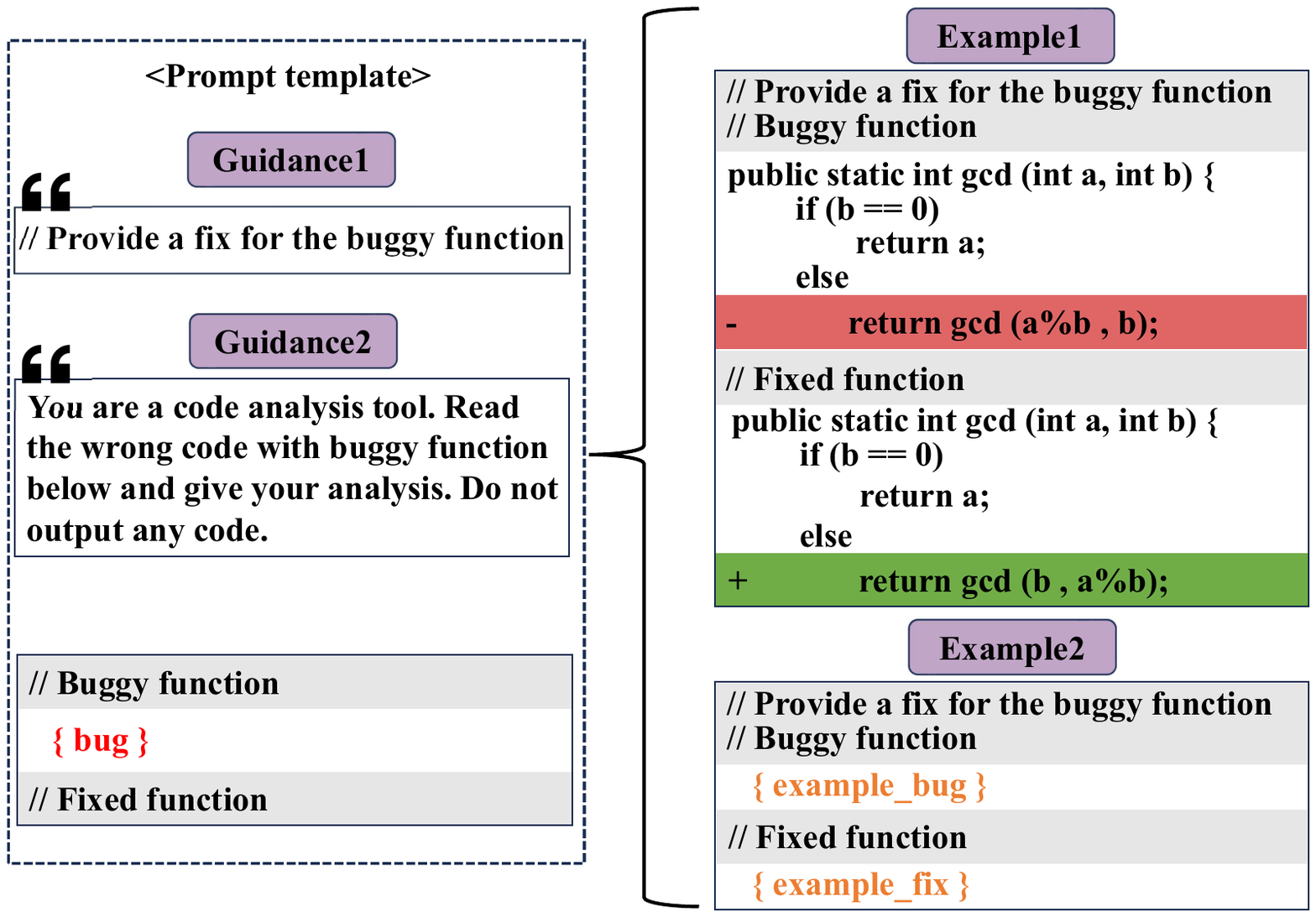} 
\caption{Prompt structure}
\label{Fig.prompt} 
\end{figure}

Figure \ref{Fig.prompt} illustrates the prompt structure utilized in this experiment. 
There is a fundamental structure that consists of three parts. First, the line ``// Buggy function'' is provided, followed by the buggy function that requires repair. Finally, the line ``// Fixed function'' is presented. For convenience, we refer to this structure as the ``Basic'' in the following sections.

In our subsequent experiments, we employed four types of prompts: 1) Two-shot prompt. Following prior work~\cite{xia2023automated,li2025bybrid}, we provide two examples. The first example, crafted by us, involves finding the greatest common divisor of two numbers. This example is adapted to match the programming language used by the buggy function under repair. For instance, if the buggy function is written in Java, our example is also implemented in Java. The second example is the shortest bug, other than the one under repair, from the same project as the buggy function. We maintain the style of the first example by providing both the erroneous version and the correct fix. Finally, we include Guidance 1, which describes the function-level bug repair task, along with the basic structure. This prompt will be used in RQ1 on the Defects4J and BugsCpp datasets. 2) One-shot prompt. Since bugs from algorithmic assignments do not have the concept of being from the same project, we remove the second example from the two-shot prompt while keeping the rest unchanged. This prompt will be used in RQ2. 3) Zero-shot prompt. To investigate whether adding examples in the prompt affects the repair performance of large models, we further remove the first example from the one-shot prompt, retaining only Guidance 1 and the basic structure. This prompt will be used in RQ4. 4) Analysis prompt. To explore whether including an analysis of the erroneous code in the prompt influences the repair performance of large models, we use DeepSeek-Coder to generate one analysis for each bug on the ConDefects-Java and ConDefects-Python datasets. To accomplish this task, we replace Guidance 1 with Guidance 2, which describes this specific task. Additionally, we remove the last line of the basic structure, requiring the large model to output only an analysis of the erroneous code, rather than the correct code. The structure and usage of the prompts are summarized in Table \ref{tab:prompt_design}.
\input{tables/prompt}

\subsection{Patch Generation and Validation}
Following prior works~\cite{xia2023automated,li2025bybrid}, we initially generate 200 patches per bug for \dfj{} and \bugscpp{} datasets. However, our preliminary study (Section~\ref{RQ:RQ2}) revealed that plausible patches predominantly appeared within the first 30 generated patches, with diminishing returns thereafter. Consequently, we reduced the generation budget to 30 patches per bug in subsequent experiments. To isolate the evaluation of patch generation capability, we provided method-level perfect fault localization, eliminating potential confounding factors introduced by localization inaccuracies.

For patch validation, we first de-duplicated all generated patches. Following existing works~\cite{xia2023automated,ouyang2024benchmarking,yang2023large}, we leverage the comprehensive test suites provided in the datasets to evaluate patch correctness: a patch that passes all test cases is considered as a \textbf{plausible patch}. All plausible patches are then manually inspected by the first two authors. If a patch is semantically equivalent to the ground truth patch, it is classified as a \textbf{correct patch}. This two-stage validation process ensures both functional correctness and semantic alignment with developer fixes. We also calculate the \textbf{Repair Rate}, defined as the ratio of bugs with correct patches to the total number of bugs in the dataset, and the \textbf{Precision} of the generated patches, defined as the ratio of correctly patched bugs to the total number of plausibly patched bugs.

%% file: tables/dataset.tex

\begin{table}[htbp]
  \caption{Overview of Evaluation Datasets}
  \label{tab:datasets}
  \centering
  \resizebox{\columnwidth}{!}{
  \begin{tabular}{p{1.5cm}lccr}
    \toprule
    \textbf{Source Type} & \textbf{Dataset Name} & \textbf{Year} & \textbf{Language} & \textbf{\# Instances} \\
    \midrule
    \multirow{3}{1.5cm}{Enterprise\\Projects} & Defects4J v1.2~\cite{just2014defects4j}       & 2014 & Java     & 255    \\
                                               & Defects4J v2.0~\cite{just2014defects4j}       & 2015 & Java     & 228   \\
                                               & BugsCpp~\cite{10298287}              & 2023 & C/C++    & 106   \\
    \midrule
    \multirow{4}{1.5cm}{Algorithmic\\Assignments} & IntroClass-C~\cite{le2015manybugs}         & 2015 & C      & 297   \\
                                                    & IntroClass-Java~\cite{durieux2016introclassjava}      & 2016 & Java   & 297   \\
                                                    & ConDefects-Java~\cite{wu2023ConDefectsnewdatasetaddress}          & 2023 & Java   & 563   \\
                                                    & ConDefects-Py~\cite{wu2023ConDefectsnewdatasetaddress}        & 2023 & Python & 563   \\
    \bottomrule
  \end{tabular}
  }
\end{table}

%% file: tables/llm.tex
\begin{table}[htbp]
    \caption{Evaluated LLMs in This Study}
    \label{tab:models}
    \centering
    \resizebox{\columnwidth}{!}{
        \begin{tabular}{lccccc}
            \toprule
            \textbf{LLM} & \textbf{Provider}& \textbf{Year} & \textbf{\#Size$\uparrow$}  & \textbf{Code-Specialized} \\
            \midrule
            \codellama{}-7B~\cite{roziere2023code} & \meta{} Meta & 2023 & 7B  & \textcolor{dimGreen}{\ding{51}} \\
            \llama{}-2-13B~\cite{touvron2023llama} & \meta{} Meta & 2023 & 13B & \textcolor{red}{\ding{55}} \\
            \starcoder{}~\cite{li2023starcoder} & \bigcode{} BigCode & 2023 & 15.5B & \textcolor{dimGreen}{\ding{51}}\\
            \deepseekcoder{}-33B-instruct~\cite{guo2024deepseekcoderlargelanguagemodel} & \deepseek{} DeepSeek  & 2024 & 33B & \textcolor{dimGreen}{\ding{51}} \\
            \bottomrule
        \end{tabular}
    }
\end{table}


%% file: tables/prompt.tex
\begin{table}[htbp]
\caption{Prompt Settings Used in This Study}
\label{tab:prompt_design}
\centering
\footnotesize
\begin{tabular}{c|c|c}
\toprule
\textbf{ID} & \textbf{Prompt Settings} & \textbf{RQ} \\
\midrule
1 & Example 1 + Example 2 + Guidance 1 + Basic & RQ1 \\
\midrule
2 & Example 1 + Guidance 1 + Basic & RQ2,RQ4 \\
\midrule
3 & Guidance 1 + Basic & RQ4 \\
\midrule
4 &Guidance 2 + ModifiedBasic & RQ4 \\
\bottomrule
\end{tabular}
\end{table}

%% file: 5ResultAnalysis.tex
\section{Results}

\subsection{RQ1: How effective are LLMs in repairing bugs from enterprise-grade projects?}
\label{RQ:RQ1}

\input{tables/table1}

\subsubsection{Results on \dfj{}}

Table~\ref{tab:table1} presents the repair results of the four selected LLMs (i.e., \codellama{}, \llama{}, \starcoder{}{}, \deepseekcoder{}). It can be observed that all four LLMs achieve significantly better performance on the \dfj{} dataset than on \bugscpp{}. Focusing on the results for \dfj{}, \deepseekcoder{} achieves the best repair performance among the four LLMs, successfully fixing 44 and 50 bugs on \dfjvone{} and v2.0, respectively. It also achieves the highest precision on both versions, with 69.8\% and 71.4\%, respectively. Although \starcoder{}{} has nearly half the number of parameters compared to \deepseekcoder{} (15.5B vs. 33B), its performance does not degrade proportionally. It successfully repairs 42 and 44 bugs on \dfjvone{} and v2.0, respectively, while maintaining a favorable precision of 60.9\% and 68.8\%. This demonstrates that \starcoder{}{} achieves a competitive balance between effectiveness and efficiency. In addition, \llama{} exhibits the weakest performance among the four LLMs, producing only 19 and 18 correct repairs on \dfjvone{} and v2.0, respectively. Its precision is also relatively low, at 30.2\% and 60.0\%. In contrast, despite having a significantly smaller parameter size (7B vs. \llama{}'s 13B), \codellama{} demonstrates substantially better repair capabilities, generating 40 and 34 correct fixes on the two datasets, respectively. Notably, it achieves a much higher precision compared to \llama{} on \dfjvone{} (61.5\% vs. 30.2\%) and comparable precision on \dfjvtwo{} (58.6\% vs. 60.0\%).

\subsubsection{Results on \bugscpp{}}

All LLMs exhibited a significant drop in performance on the \bugscpp{} dataset, with the average RRate decreasing from 15.1\% on \dfj{} to just 3.5\% — a reduction of nearly 76.5\%. Although \deepseekcoder{} still achieves the best repair performance among the four LLMs, it correctly fixes only 6 bugs on \bugscpp{}. Consistent with the results on \dfj{}, \codellama{} continues to outperform \llama{}, despite having fewer parameters. 
\begin{figure}[htbp] 
\centering 
\includegraphics[width=\columnwidth,trim=2cm 6.5cm 2cm 5.5cm,clip]{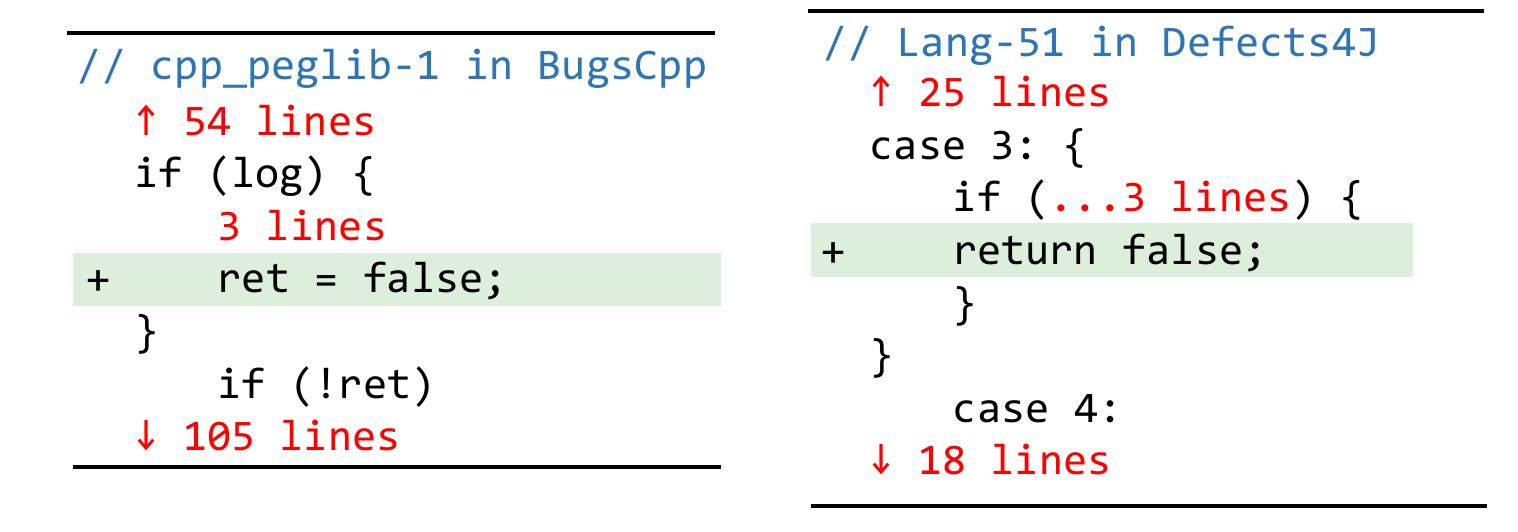} 
\caption{A bug on \bugscpp{} similar to \dfj{}} 
\label{Fig.exmaple1} 
\end{figure}

Moreover, we analyzed the reasons behind the sharp drop in LLMs' repair performance on the \bugscpp{} dataset. This drop can be attributed to the limited cross-language generalizability of LLMs. As shown in Figure~\ref{Fig.exmaple1}, the bug in \bugscpp{} are actually quite similar to Lang-51 in \dfj{}. However, all four LLMs failed to fix this bug due to its substantial length. The bug spans 166 lines, which is more than three times the length of Lang-51 (51 lines), posing significant challenges for large models in bug localization. On \deepseekcoder{}, a large proportion of the generated patches either fail to compile (20\%) or modify incorrect locations (76\%), ultimately resulting in unsuccessful repairs. This phenomenon has garnered our attention, and we will conduct further analysis of bug length in RQ3(\ref{RQ:RQ3}).

\begin{findingbox}
\textbf{Finding 1:} The LLMs achieved satisfactory repair performance on the Defects4J dataset, but their effectiveness dropped significantly on the BugsCPP dataset. This indicates that the generalizability of large language models for repairing bugs in enterprise-grade projects is not satisfactory.
\end{findingbox}

\begin{findingbox}
\textbf{Finding 2:} Fine-tuning on code-related tasks can enhance the repair performance of large models. However, an increase in model parameter size does not necessarily lead to a proportional improvement in repair effectiveness.
\end{findingbox}

\subsubsection{Complementary Analysis}

\begin{figure}[tbp] 
\centering 
\includegraphics[width=\columnwidth,trim=0cm 4.8cm 0cm 4cm,clip]{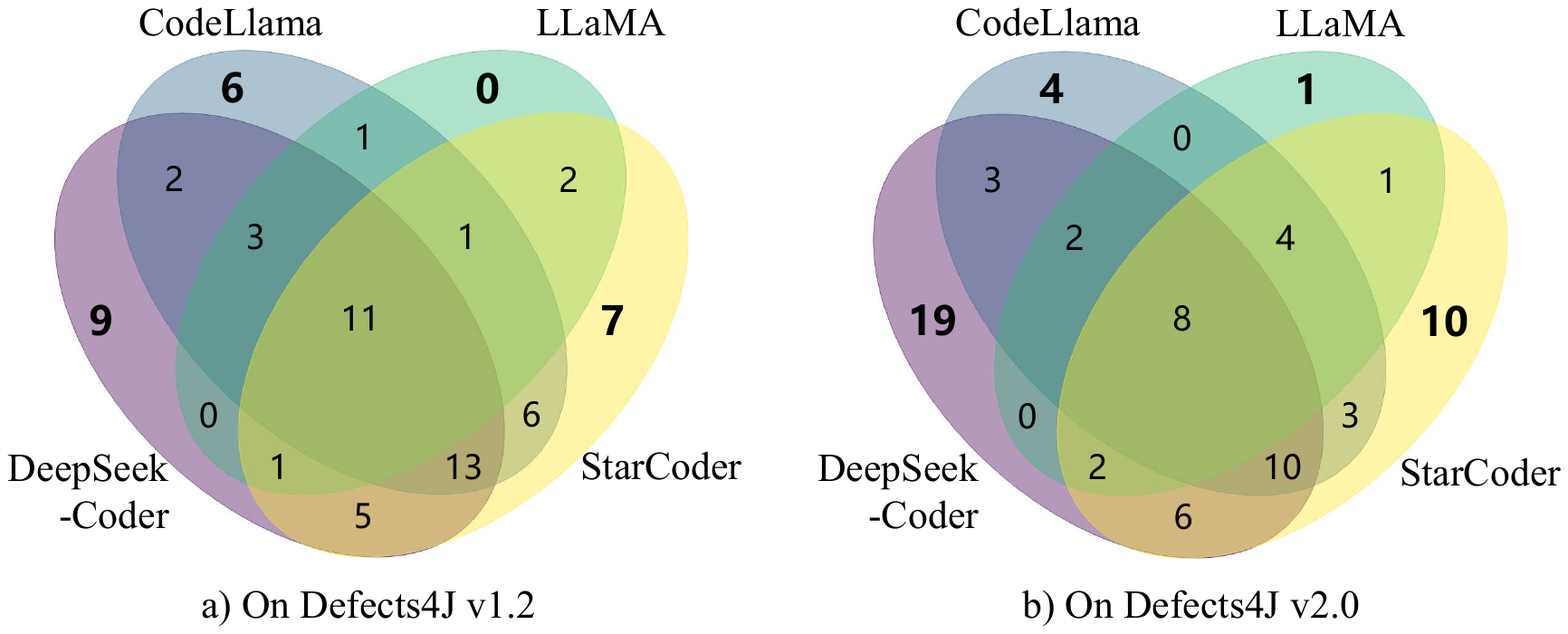} 
\caption{Unique bugs fixed by LLMs on \dfjvone{} \& v2.0} 
\label{Fig.d4j} 
\end{figure}

As illustrated in Figure~\ref{Fig.d4j}, the fine-tuned \deepseekcoder{} model achieved 50 correct fixes on \dfjvtwo{}, with 19 being unique repairs unobtainable by the other three LLMs, further demonstrating its exceptional repair capabilities. \starcoder{} ranked second, generating 44 correct fixes including 10 unique ones, while \codellama{} produced 34 correct fixes (4 unique). Notably, despite its substantially lower repair count compared to code-specialized models, the non-fine-tuned \llama{} still contributed 1 unique repair. Similar results were observed on the \dfjvone{}. Critically, all four evaluated LLMs exhibited complementary repair capabilities, with each generating specific fixes unattainable by others. This observation highlights the potential benefits of ensemble or hybrid approaches in practical APR workflows to maximize repair coverage.

\begin{findingbox}
\textbf{Finding 3:} The patches generated by LLMs exhibit complementarity, suggesting that employing multiple large models in repair practices can enhance repair effectiveness.
\end{findingbox}

\subsection{RQ2: How effective are LLMs in repairing bugs in algorithmic assignments?}
\label{RQ:RQ2}

\begin{figure}[h]
  \centering

  \subfigure[Four LLMs on \dfj{}]{
     \includegraphics[width=\linewidth]{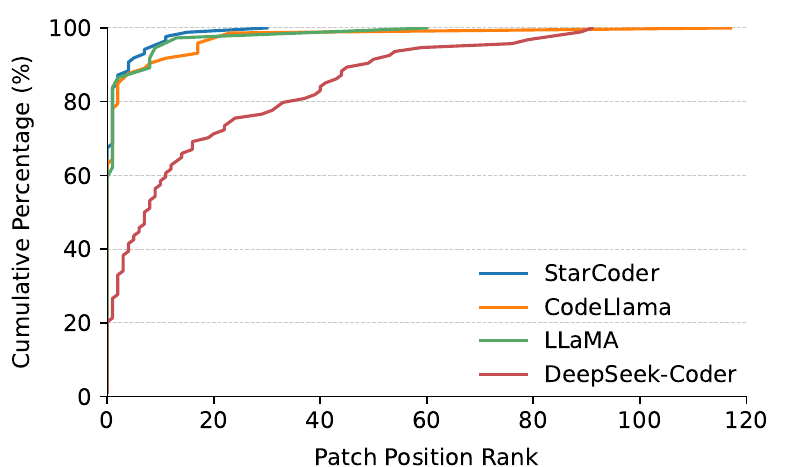}
     \label{Fig.patch}
  }
 \subfigure[\starcoder{} on \introclassc{} \& \introclassjava{}]{
     \includegraphics[width=\linewidth]{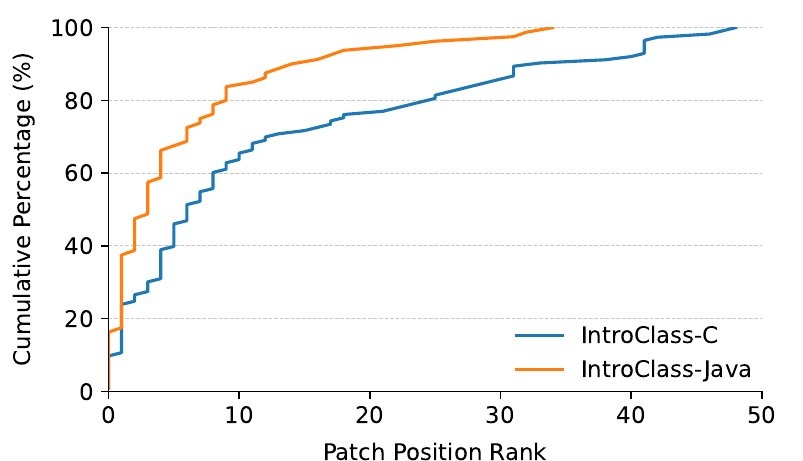}
     \label{Fig.intro}
  }
  \caption{Distribution of Correct Patch Rankings} 

  \label{Fig.pr}
\end{figure}

In RQ1(\ref{RQ:RQ1}), each LLM was configured to generate 200 patches per bug on the \dfj{} and \bugscpp{} datasets. However, during manual verification of the generated patches, we observed that the vast majority of correct patches tended to appear among the top candidates. To investigate this further, we analyzed the rank positions at which the first correct patch appeared. As shown in Figure~\ref{Fig.patch}, we found that most correct patches emerged within the top 30 candidates. There are several notable data points in the figure. Among the correct patches generated by \llama{}, only one appears at rank 61, while the rest are within the top 15. For the correct patches generated by \starcoder{}, only one patch is ranked at the 31st position, with the rest located within the top 30. In the case of \codellama{}, only one correct patch appears at rank 118, while the rest are within the top 25. In terms of proportion, 76.59\% of the correct patches generated by \deepseekcoder{} are within the top 30. In contrast, the other large models each have only one correct patch that is not within the top 30. To further validate this finding, we conducted experiments using \starcoder{} on both the \introclassc{} and \introclassjava{} datasets, with the results presented in Figure~\ref{Fig.intro}. Consistent with our previous observations, 89.6\% of the correct patches were found within the top 30 positions. Based on this observation, we limited the number of generated patches to 30 per bug, which allowed us to retain most of the repair effectiveness while significantly reducing resource consumption.

\begin{findingbox}
\textbf{Finding 4:} Increasing the number of candidate patches shows diminishing returns in terms of repair effectiveness. In our experiments, generating 30 candidate patches achieves a better balance between repair success rate and computational cost.
\end{findingbox}

\input{tables/table2}

\subsubsection{Overall performance}
Table \ref{tab:table2} shows the repair results of the four selected LLMs in fixing bugs from algorithmic assignments. While \codellama{}, \starcoder{}, and \deepseekcoder{} perform better on algorithmic assignment bugs than on enterprise-grade projects, \llama{} shows a further decline in performance in this setting. Specifically, \deepseekcoder{} consistently achieves the best repair performance among the four selected LLMs, achieving the highest number of correctly fixed bugs across all evaluation datasets. In contrast, \llama{} generates the fewest correct patches on every dataset. The strong scaling effect described in Finding 2 remains evident among the three code-specialized LLMs.

\subsubsection{Performance Across Programming Languages}
\begin{itemize}
    \item  \textbf{C vs. Java.} By examining the performance of the four LLMs on the \introclassc{} and \introclassjava{} datasets, we observe a clear difference in repair results across programming languages. For C programs, LLMs tend to generate more plausible and correct fixes at the cost of lower precision. In contrast, for Java programs, LLMs consistently achieve exceptionally high precision, with all LLMs exceeding 98\%, which is particularly impressive given the challenging nature of achieving such high precision in the APR domain. This indicates that the selected LLMs tend to generate more reliable repairs for Java programs in our evaluated datasets, which aligns with the results in RQ1 (Section~\ref{RQ:RQ1}) where the precision for \dfj{} was consistently higher than that for \bugscpp{}. As can be seen from the previous experiments (Fig~\ref{Fig.intro}), the ranking of correct patches on the Java dataset is higher than that on the C dataset, which further corroborates our findings.
    \item  \textbf{Python vs. Java.} On the \condefectspython{} and \condefectsjava{} datasets, we did not observe significant differences in the overall repair effectiveness of the LLMs. While \codellama{} and \deepseekcoder{} generated more correct and plausible patches on \condefectsjava{}, their precision was lower than on \condefectspython{}. In contrast, \llama{} and \starcoder{} generated fewer correct and plausible patches on \condefectsjava{}, but achieved higher precision. This suggests that the evaluated LLMs exhibit varying trade-offs between recall and precision depending on the programming languages.
\end{itemize}

\begin{findingbox}
\textbf{Finding 5:} Syntactic variations in semantically equivalent code bugs lead to divergent repair outcomes. This demonstrates that coding style substantially influences the repair performance of LLMs, even when program semantics remain invariant.
\end{findingbox}

\subsection{RQ3: Effects of Bug Length and Patch Actions}
\label{RQ:RQ3}

\begin{figure}[tbp] 
\centering 
\includegraphics[width=0.45\textwidth]{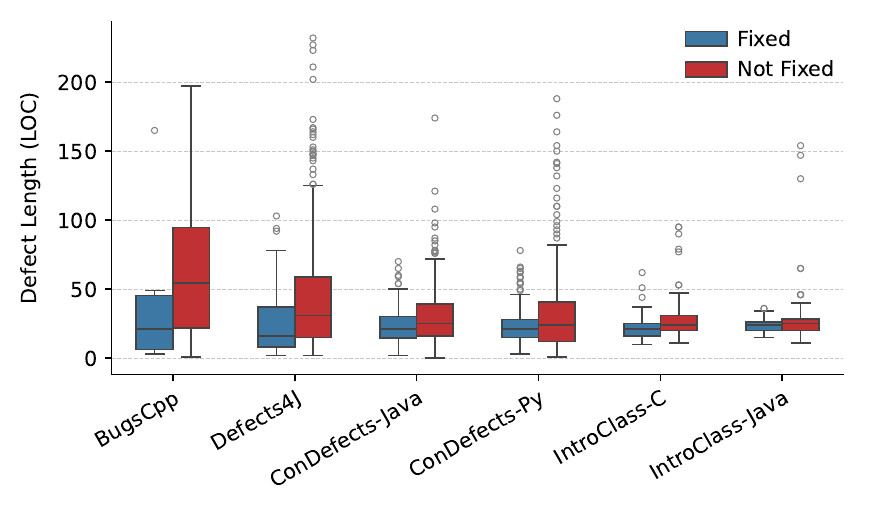} 
\caption{Length distribution of successfully repaired and unrepaired bugs}

\label{Fig.main2} 
\end{figure}

\subsubsection{length}
We collected statistics on the lengths of successfully and unsuccessfully repaired bugs by the four selected LLMs across all six datasets. As shown in Figure~\ref{Fig.main2}, successfully repaired bugs tend to be significantly shorter than those that failed to be repaired. This trend is clearly reflected in the median values, where the median length of successfully repaired bugs is consistently lower than that of unsuccessful ones. Moreover, LLMs exhibit pronounced difficulties in repairing long functions, especially those exceeding 100 lines, where their effectiveness drops significantly.

\begin{findingbox}
\textbf{Finding 6:} Bugs of shorter length are more likely to be successfully repaired by LLMs. This trend remains consistent across programming languages, highlighting current limitations of LLMs in addressing longer or more complex bugs.
\end{findingbox}

\subsubsection{repair actions}

\begin{figure*}[h]
  \centering

  \subfigure[Java top 10 code operations]{
     \includegraphics[width=0.31\linewidth]{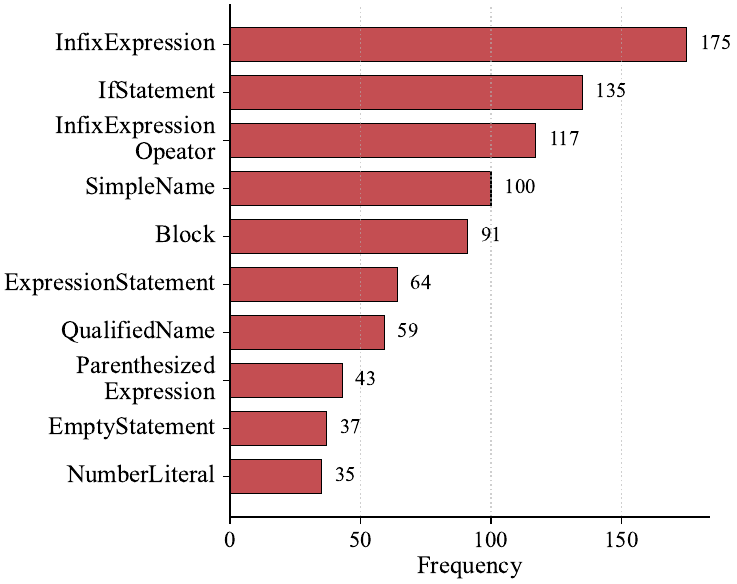}
     \label{Fig.java}
  }
 \subfigure[C top 10 code operations]{
     \includegraphics[width=0.31\linewidth]{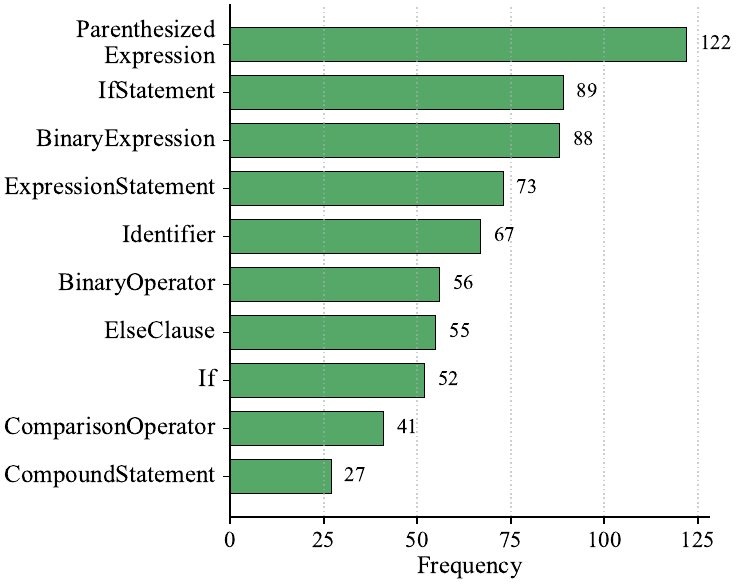}
     \label{Fig.c}
  }
  \subfigure[Python top 10 code operations]{
     \includegraphics[width=0.31\linewidth]{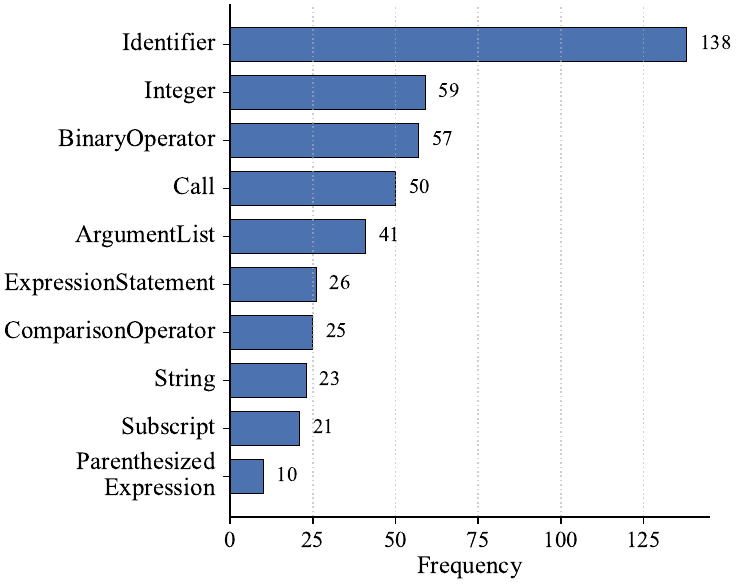}
     \label{Fig.python}
  }
  \caption{Top 10 modification operations in Java, C, and Python} 

  \label{Fig.actions}
\end{figure*}

We selected all correctly repaired bugs and used GumTree~\cite{falleri2014fine,martinez2023hyperparameter} to extract the edit operations between the original buggy code and the corresponding repair patches. For each programming language, we analyzed the distribution of these operations to identify which code elements were most frequently modified. The results of this analysis are presented in Figure~\ref{Fig.actions}.

For Java, the most frequent code changes involve \codeIn{InfixExpression} and \codeIn{IfStatement}, with occurrences of 175 and 135, respectively.

The frequent appearance of \codeIn{QualifiedName} and \codeIn{ExpressionStatement} further highlights the importance of name resolution and expression evaluation in Java-related repairs. 
In the C dataset, the most common edit operations include \codeIn{ParenthesizedExpression}, \codeIn{BinaryExpression}, and \codeIn{IfStatement} each occurring over 90 times. This indicates that repairs in C often center around complex expressions and conditional constructs. The high frequency of \codeIn{BinaryOperator} and \codeIn{ComparisonOperator} also indicates that arithmetic and logical operations are frequent sources of bugs. 
For Python, \codeIn{Identifier} is the most commonly modified element, appearing 138 times, followed by \codeIn{Integer} and \codeIn{BinaryOperator}. This indicates that variable usage and numerical computations are central to bug patterns in Python. The notable frequency of \codeIn{Call} and \codeIn{ArgumentList} edits suggests that function invocations and parameter handling are common areas of bugs.

\begin{findingbox}
\textbf{Finding 7:} The distinct characteristics of different programming languages, in terms of the operations required for repair, also have an impact on the repair performance of large models.
\end{findingbox}

\subsection{RQ4: Impact of Prompt Settings on Repair Performance}
\label{RQ:RQ4}

\subsubsection{one-shot vs zero-shot}

\input{tables/table4} 

As shown in Table~\ref{tab:table4}, all LLMs exhibit reduced performance in the zero-shot setting compared to the one-shot setting, indicating that the presence of in-context example substantially enhances their repair effectiveness.

Focusing on the \condefectsjava{} dataset, the average RRate of LLMs in the zero-shot setting is 22.9\% lower than in the one-shot setting, dropping from 11.5\% to 8.9\%. Specifically, \llama{} exhibits the most substantial drop in correct repairs, decreasing by 85.7\% and successfully fixing only 1 bug in the zero-shot setting. The number of correct repairs generated by \codellama{} and \starcoder{} decreased by 44\% and 29.3\%, respectively. \deepseekcoder{} exhibits the smallest reduction at 7.1\%, yet still shows a notable decrease. In terms of precision, it also declines from 74.4\% in the one-shot setting to 71.4\% in the zero-shot setting.
On the \condefectspython{} dataset, a similar downward trend is also observed. All four LLMs exhibit a decrease in the number of correctly repaired bugs under the zero-shot setting. \llama{} again has the largest drop at 77.8\%, falling from 9 to 2 correct repairs. \codellama{}'s decline of 67.4\% is even more significant than its drop on the \condefectsjava{} dataset. \starcoder{}'s number of correct repairs falls by 31.6\%, while \deepseekcoder{}'s experiences the smallest decrease at 23.1\%. Repair precision also shows a consistent downward trend across all LLMs.

\begin{findingbox}
\textbf{Finding 8:} Providing examples of repair tasks in prompts substantially improves the repair performance of LLMs, with varying degrees of improvement across different LLMs.
\end{findingbox}

\subsubsection{Bug Analysis}

\input{tables/table5}

The experimental results in Table~\ref{tab:table4} reveal a striking contrast in how bug analysis integration affects LLMs with varying repair capabilities. For weaker models like \codellama{} and \llama{}, incorporating bug analysis leads to substantial gains: \codellama{} achieves a 53.6\% increase in correct patches on \condefectsjava{} (from 28 to 43) and a remarkable 300\% increase on \condefectspython{} (from 15 to 60). \llama{} shows the most significant improvements, with correct repairs rising from 1 to 32 (a 3100\% increase) on \condefectsjava{} and from 2 to 32 (1500\%) on \condefectspython{}.

In contrast, stronger models like \deepseekcoder{} experience performance degradation when using bug analysis-augmented prompts: the number of correct repairs drops by 46.6\% on \condefectsjava{} (from 118 to 63) and by 22.6\% on \condefectspython{} (from 93 to 72). This decline is likely due to conflicts between the model's advanced reasoning capabilities and potentially inaccurate diagnostic cues in the provided analysis. 

\begin{figure}[htbp] 
\centering 
\includegraphics[width=0.49\textwidth,trim=0cm 4cm 2cm 4cm,clip]{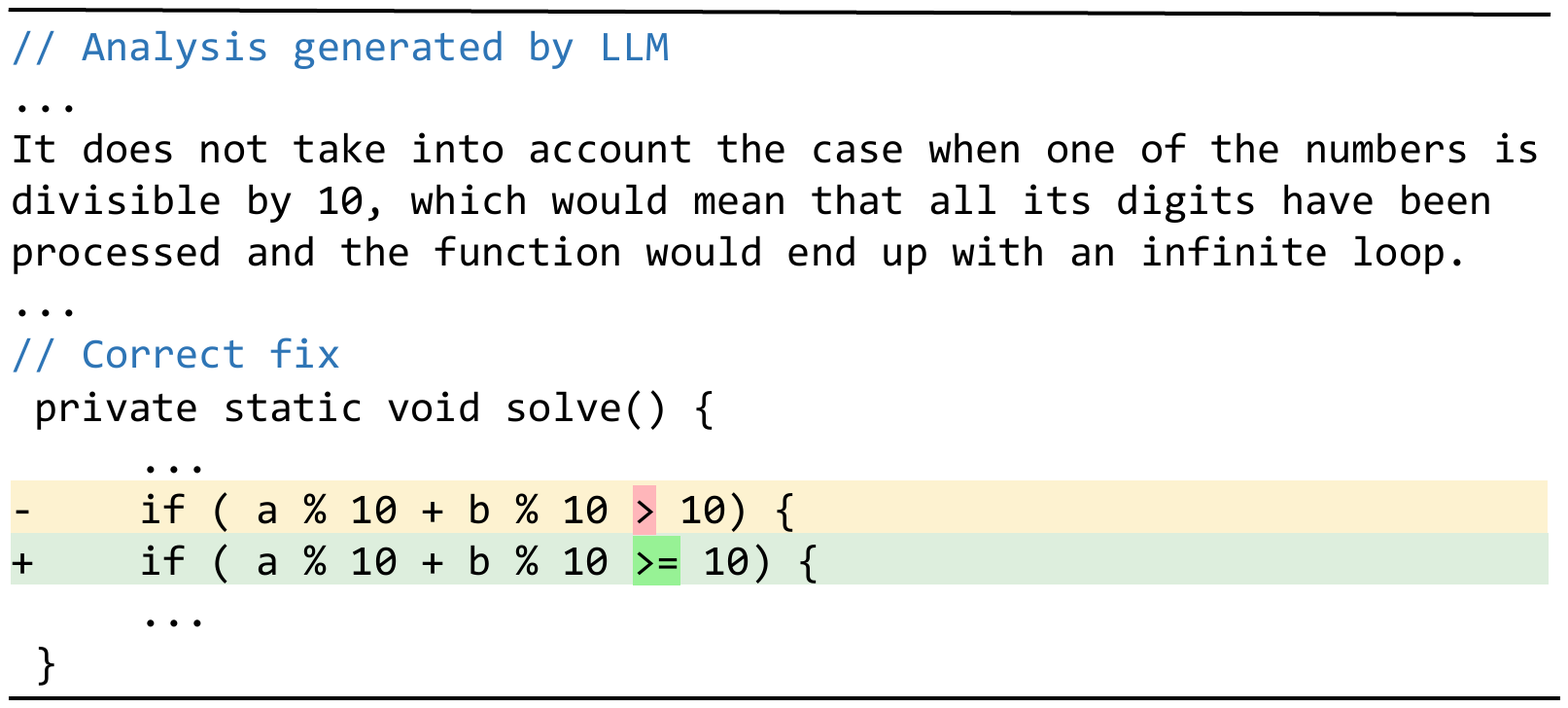} 
\caption{An example bug abc229-b from \condefectsjava{} dataset}
\label{Fig.abc229_b} 
\end{figure}

This counterintuitive phenomenon warrants further investigation. Through manual inspection, we identified that the performance drop is primarily caused by flawed diagnostic information in analysis. Figure \ref{Fig.abc229_b} presents an example. In this case, the correct fix is to change the condition from ``$>$" to ``$>$=" in the code.  When the initial analysis misidentifies the root cause, the LLMs tend to follow the incorrect reasoning path in the prompt, resulting in incorrect repairs. This can explain why models with poorer performance (\llama{} and \codellama{}) increased the number of repairs when provided with analysis, while models with better performance showed the opposite trend. On the one hand, analysis generated by LLMs may be wrong, which can mislead models, preventing them from completing repairs. Therefore, models that successfully repaired more bugs without analysis are more susceptible to the negative impact of incorrect analysis. On the other hand, correct analysis can assist large models in completing repairs. However, models with stronger performance had already completed these repairs without analysis, and thus they gained less benefit from the analysis. Consequently, the above results are presented. This observation reveals the strong dependency of LLMs on prompt content in APR and highlights the importance of ensuring the accuracy of repair-related information provided in the prompt for real-world applications.

\begin{figure}[htbp] 
\centering
\includegraphics[width=\columnwidth,trim=0cm 2.5cm 0cm 3.6cm,clip]{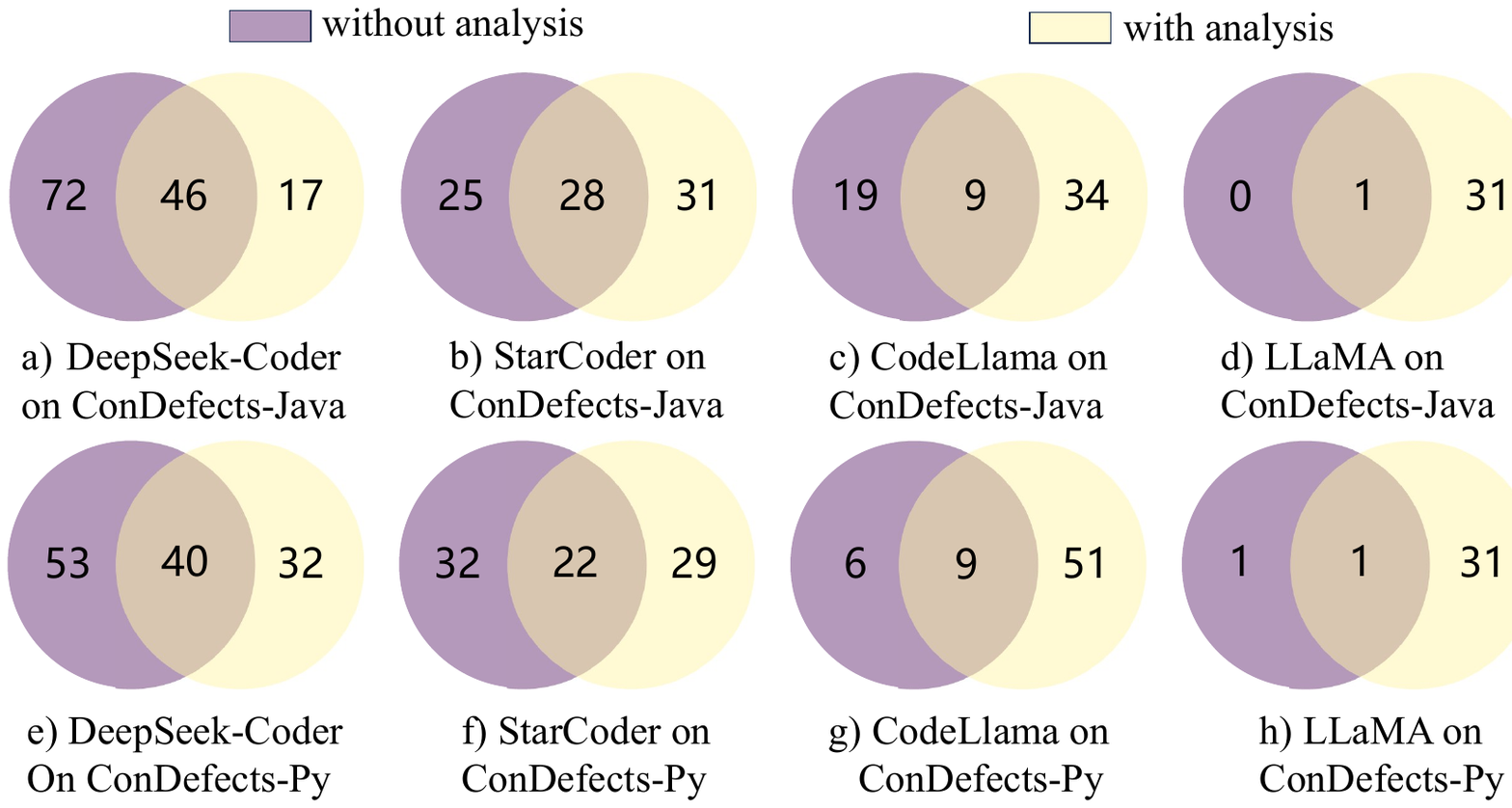}
\caption{Unique bugs fixed by LLMs with and without Bug Analysis}

\label{Fig.main3} 
\end{figure}

Although the bug analysis may contain inaccuracies, incorporating them still enables the repair of bugs that could not be fixed under zero-shot setting. As shown in Figure~\ref{Fig.main3}, all four selected LLMs are able to repair previous unresolved bugs across both datasets. This demonstrates that applying LLMs in a chain-of-thought (COT) manner (first provide bug analysis and then generate patches) can yield measurable improvements over the single-stage approaches, highlighting the complementary value of analytical reasoning in APR systems.

\begin{findingbox}
\textbf{Finding 9:} LLMs exhibit significant dependency on the contextual information provided in prompts for APR. Notably, incorporating error analysis into the prompts demonstrably enhances repair effectiveness.
\end{findingbox}

%% file: tables/table1.tex

\begin{table}[htbp]
\centering
\caption{Repair results of LLMs on bugs from enterprise-grade projects. \textit{In this table, C/P denotes the number of correct fixes over the number of plausible fixes, RRate denotes the repair rate, and Prec denotes precision.}}
\resizebox{\columnwidth}{!}{
\begin{tabular}{l|c|c|c|c|c}
\toprule
\textbf{Dataset} & \textbf{Metrics} & \textbf{\codellama{}}  & \textbf{\llama{}} & \textbf{\starcoder{}} & \textbf{\deepseekcoder{}}\\
\midrule
\multirow{3}{*}{\shortstack{\textbf{\dfjvone{}} \\ (255 bugs)}}
& C/P &  40/65 & 19/63 & 42/69 & 44/63 \\
& RRate (\%)  & 15.7 & 7.5 & 16.5 & 17.3 \\
& Prec (\%)    & 61.5 & 30.2 & 60.9 & 69.8 \\
\midrule
\multirow{3}{*}{\shortstack{\textbf{\dfjvtwo{}} \\ (228 bugs)}}
& C/P & 34/58 & 18/30 & 44/64 & 50/70 \\
& RRate (\%)  & 14.9  & 7.9   & 19.3  & 21.9 \\
& Prec (\%)    & 58.6  & 60    & 68.8 & 71.4 \\
\midrule
\multirow{3}{*}{\shortstack{\textbf{\bugscpp{}} \\ (106 bugs)}}
& C/P & 5/20 & 1/17 & 3/23 & 6/27 \\
& RRate (\%)  & 4.7  & 0.9  & 2.8  & 5.7 \\
& Prec (\%)    & 25.0  & 5.9 & 13.0 & 22.2 \\
\bottomrule
\end{tabular}
}
\label{tab:table1}
\end{table}

%% file: tables/table2.tex

\begin{table}[htbp]
\centering
\caption{Repair results of LLMs on bugs from algorithmic assignments. \textit{In this table, C/P denotes the number of correct fixes over the number of plausible fixes, RRate denotes the repair rate, and Prec denotes precision.}}
\resizebox{\columnwidth}{!}{
\begin{tabular}{l|c|c|c|c|c}
\toprule
\textbf{Dataset} & \textbf{Metrics} & \textbf{\codellama{}}  & \textbf{\llama{}} & \textbf{\starcoder{}} & \textbf{\deepseekcoder{}}\\
\midrule
\multirow{3}{*}{\shortstack{\textbf{\introclassc{}} \\ (297 bugs)}}
& C/P & 59/65 & 16/17 & 72/77 & 135/143 \\
& RRate (\%)  & 19.9 & 5.4 & 24.2 & 45.5 \\
& Prec (\%)    & 90.8 & 94.1 & 93.5 & 94.4 \\
\midrule
\multirow{3}{*}{\shortstack{\textbf{\introclassjava{}} \\ (297 bugs)}}
& C/P & 51/52 & 5/5 & 68/69 & 125/127 \\
& RRate (\%)  & 17.2  & 1.7 & 22.9  & 42.1 \\
& Prec (\%)    & 98.1  & 100 & 98.6  & 98.4 \\
\midrule
\multirow{3}{*}{\shortstack{\textbf{\condefectsjava{}} \\ (563 bugs)}}
& C/P & 50/73 & 7/9  & 75/96 & 127/170 \\
& RRate (\%)  & 8.9   & 1.2  & 13.3 & 22.6 \\
& Prec (\%)    & 68.5  & 77.8 & 78.1 & 74.7 \\
\midrule
\multirow{3}{*}{\shortstack{\textbf{\condefectspython{}} \\ (563 bugs)}}
& C/P & 46/63  & 9/10 & 79/103  & 121/153 \\
& RRate (\%)  & 8.2    & 1.6  & 14.0   & 21.5 \\
& Prec (\%)    & 73.0  & 90.0 & 76.7 & 79.1 \\
\bottomrule
\end{tabular}
}
\label{tab:table2}
\end{table}

%% file: tables/table4.tex
\begin{table*}[ht]
\centering
\caption{Repair results of LLMs on \condefectsjava{} and \condefectspython{} with different prompt settings. \textit{In this table, C/P denotes the number of correct fixes over the number of plausible fixes, RRate denotes the repair rate, and Prec denotes precision.}}
\resizebox{\textwidth}{!}{
\begin{tabular}{l|l|ccc|ccc|ccc|ccc}
\toprule
\multirow{2}{*}{Dataset} & \multirow{2}{*}{Prompt} 
& \multicolumn{3}{c|}{\codellama{}} 
& \multicolumn{3}{c|}{\llama{}} 
& \multicolumn{3}{c|}{\starcoder{}} 
& \multicolumn{3}{c}{\deepseekcoder{}} \\
& & C/P & RRate (\%) & Prec (\%) & C/P & RRate (\%) & Prec (\%) & C/P & RRate (\%) & Prec (\%) & C/P & RRate (\%) & Prec (\%) \\
\midrule

\multirow{3}{*}{\shortstack{\textbf{\condefectsjava{}} \\ (563 bugs)}} 
& zero-shot        & 28/39 & 5.0  & 71.8 & 1/3    & 0.2 & 33.3   & 53/81 & 9.4    & 65.4   & 118/157 & 21.0   & 75.2   \\
& one-shot         & 50/73 & 8.9  & 68.5 & 7/9    & 1.2 & 77.8   & 75/96 & 13.3   & 78.1   & 127/170 & 22.6   & 74.7   \\
& bug analysis   & 43/72 & 7.6  & 59.2 & 32/52  & 5.7 & 61.5   & 59/75 & 10.5   & 78.7   & 63/95   & 11.2   & 66.3   \\

\midrule

\multirow{3}{*}{\shortstack{\textbf{\condefectspython{}} \\ (563 bugs)}} 
& zero-shot        & 15/25   & 2.7    & 60.0   & 2/3     & 0.4    & 66.7   & 54/89   & 9.6    & 60.7    & 93/149   & 16.5   & 62.4   \\
& one-shot         & 46/63   & 8.2    & 73.0   & 9/10    & 1.6    & 90.0   & 79/103  & 14.0   & 76.7    & 121/153  & 21.5   & 79.1   \\
& bug analysis  & 60/79   & 10.7   & 75.9   & 32/44   & 5.7    & 72.7   & 51/79   & 9.1    & 64.6    & 72/104   & 12.8   & 69.2   \\

\bottomrule
\end{tabular}
}
\label{tab:table4}
\end{table*}

%% file: tables/table5.tex

%% file: 6Discussion.tex
\section{Discussion}

\subsection{Data Leakage}
In empirical studies involving LLMs, data leakage is a critical concern. Since most LLMs are trained on large-scale code corpora mined from GitHub or other public sources, it is plausible that some of the buggy or fixed code used in selected benchmarks may partially or fully exist in the model's training data. To mitigate this risk, we deliberately selected only open-source LLMs with varying architectures and parameter sizes from different providers. We also evaluated them on multiple datasets that differ in programming language, collection time, and source. This diversity helps to limit the influence of memorized examples on the evaluation results. Moreover, our findings focus more on the relative performance differences across programming languages, prompt settings, and LLMs, which remain informative even in the presence of some potential leakage.

\subsection{Limitation}
In our study, we selected four open-source LLMs -- \codellama{}, \llama{}, \starcoder{}, and \deepseekcoder{} -- encompassing both general-purpose models and code-specialized models, with parameter sizes ranging from 7B to 33B. However, given the rapid advancement in the development of LLMs, the capabilities of even larger or more recent models remain unexplored in our evaluation. Additionally, we employed six bug datasets covering three programming languages to analyze the performance of LLMs across multiple languages. In practice, however, real-world bugs may be more complex than those in the datasets, and additional programming languages not included in this study may pose unique challenges.  As such, further research is needed to assess LLM repair performance in broader and more complex scenarios.

\subsection{Threats to Validity}
\textbf{Internal.} We share the same threats with existing APR studies and methods. In our experiments, we manually verified each plausible patch and identified those that are semantically consistent with the developer patches as correct. Following the approach of prior work, the first two authors conducted a comprehensive manual analysis of all plausible patches and have released the full set of experimental results publicly.

`\textbf{External.} Although our study evaluates LLMs on six bug datasets spanning three programming languages, the generalizability of our findings to other programming languages, dataset types, or real-world scenarios remains an open question. Future work is needed to assess the performance of LLMs on a broader range of software bugs and languages, especially those with different syntactic or semantic characteristics.

\subsection{Implication and Future Work}

Building on our findings, we highlight two key directions for future work:

\textbf{Building APR-Specialized LLMs.} Our findings suggest that building LLMs specifically tailored for APR is a necessary direction. In our experiments, we observed that LLMs fine-tuned on code-related tasks significantly outperform their non-fine-tuned counterparts on APR benchmarks. This performance gap highlights the critical role of domain adaptation. Even the best-performing model in our evaluation, \deepseekcoder{}, demonstrated unsatisfactory results on several datasets (e.g., \bugscpp{}), indicating that there remains substantial room for improvement. Future research should explore how to construct more effective APR-specialized LLMs, including the design of training objectives, model architectures, and the use of repair-specific datasets that better reflect the challenges of real-world project bugs.

\textbf{Reducing Inference Costs.} With the continuous development of LLMs, increasing parameter sizes lead to greater computational resources and longer inference times. Our experiments indicate that generating 30 patches with an LLM can achieve performance comparable to generating 200 patches, suggesting that excessive generation may be inefficient. Furthermore, doubling the model size does not necessarily yield proportional performance gains. Notably, the smallest \codellama{} was able to produce unique repairs on evaluated datasets, highlighting the potential of smaller, more efficient models. Future research should prioritize methods to reduce inference cost, such as model compression, distillation, or efficient sampling strategies, while maintaining or improving repair accuracy. Striking a balance between performance and computational expense is critical for enabling practical, real-world adoption of APR systems to effectively support developers in fixing software bugs.

%% file: 8Conclusion.tex
\section{Conclusion}

In this study, we empirically evaluated the repair performance of four open-source LLMs across six bug datasets that span three programming languages (Java, C/C++, and Python) and two programming scenarios (bugs from enterprise projects and algorithmic assignments). By comparing the performance of LLMs on selected benchmarks, we found that the models consistently performed better on Java datasets than on C/C++ datasets. Additionally, the models performed significantly better on bugs from algorithmic assignments than those from enterprise projects. Interestingly, we observed that correct patches generated by LLMs often appear near the top of the output list, enabling a better balance between repair performance and computational overhead. In addition, LLMs’ repair effectiveness is highly sensitive to different prompt settings, and providing repair examples and analysis of the erroneous code in the prompt can guide LLM to perform better on APR task. We conducted a quantitative analysis of these factors to provide deeper insights for future research on leveraging LLMs for APR.

%% file: IEEE-conference-template-062824.bbl
\begin{thebibliography}{10}
\providecommand{\url}[1]{#1}
\csname url@samestyle\endcsname
\providecommand{\newblock}{\relax}
\providecommand{\bibinfo}[2]{#2}
\providecommand{\BIBentrySTDinterwordspacing}{\spaceskip=0pt\relax}
\providecommand{\BIBentryALTinterwordstretchfactor}{4}
\providecommand{\BIBentryALTinterwordspacing}{\spaceskip=\fontdimen2\font plus
\BIBentryALTinterwordstretchfactor\fontdimen3\font minus \fontdimen4\font\relax}
\providecommand{\BIBforeignlanguage}[2]{{%
\expandafter\ifx\csname l@#1\endcsname\relax
\typeout{** WARNING: IEEEtran.bst: No hyphenation pattern has been}%
\typeout{** loaded for the language `#1'. Using the pattern for}%
\typeout{** the default language instead.}%
\else
\language=\csname l@#1\endcsname
\fi
#2}}
\providecommand{\BIBdecl}{\relax}
\BIBdecl

\bibitem{o2017debugging}
D.~H. O'Dell, ``The debugging mindset: Understanding the psychology of learning strategies leads to effective problem-solving skills.'' \emph{Queue}, vol.~15, no.~1, pp. 71--90, 2017.

\bibitem{le2011genprog}
C.~Le~Goues, T.~Nguyen, S.~Forrest, and W.~Weimer, ``Genprog: A generic method for automatic software repair,'' \emph{Ieee transactions on software engineering}, vol.~38, no.~1, pp. 54--72, 2011.

\bibitem{xuan2016nopol}
J.~Xuan, M.~Martinez, F.~Demarco, M.~Clement, S.~L. Marcote, T.~Durieux, D.~Le~Berre, and M.~Monperrus, ``Nopol: Automatic repair of conditional statement bugs in java programs,'' \emph{IEEE Transactions on Software Engineering}, vol.~43, no.~1, pp. 34--55, 2016.

\bibitem{le2017s3}
X.-B.~D. Le, D.-H. Chu, D.~Lo, C.~Le~Goues, and W.~Visser, ``S3: syntax-and semantic-guided repair synthesis via programming by examples,'' in \emph{Proceedings of the 2017 11th Joint Meeting on Foundations of Software Engineering}, 2017, pp. 593--604.

\bibitem{long2015staged}
F.~Long and M.~Rinard, ``Staged program repair with condition synthesis,'' in \emph{Proceedings of the 2015 10th Joint Meeting on Foundations of Software Engineering}, 2015, pp. 166--178.

\bibitem{mechtaev2016angelix}
S.~Mechtaev, J.~Yi, and A.~Roychoudhury, ``Angelix: Scalable multiline program patch synthesis via symbolic analysis,'' in \emph{Proceedings of the 38th international conference on software engineering}, 2016, pp. 691--701.

\bibitem{martinez2016astor}
M.~Martinez and M.~Monperrus, ``Astor: A program repair library for java,'' in \emph{Proceedings of the 25th international symposium on software testing and analysis}, 2016, pp. 441--444.

\bibitem{Le2016HistoryDP}
\BIBentryALTinterwordspacing
X.-B.~D. Le, D.~Lo, and C.~L. Goues, ``History driven program repair,'' \emph{2016 IEEE 23rd International Conference on Software Analysis, Evolution, and Reengineering (SANER)}, vol.~1, pp. 213--224, 2016. [Online]. Available: \url{https://api.semanticscholar.org/CorpusID:8844190}
\BIBentrySTDinterwordspacing

\bibitem{long2016automatic}
F.~Long and M.~Rinard, ``Automatic patch generation by learning correct code,'' in \emph{Proceedings of the 43rd Annual ACM SIGPLAN-SIGACT Symposium on Principles of Programming Languages}, 2016, pp. 298--312.

\bibitem{xin2017leveraging}
Q.~Xin and S.~P. Reiss, ``Leveraging syntax-related code for automated program repair,'' in \emph{2017 32nd IEEE/ACM International Conference on Automated Software Engineering (ASE)}.\hskip 1em plus 0.5em minus 0.4em\relax IEEE, 2017, pp. 660--670.

\bibitem{xiong2017precise}
Y.~Xiong, J.~Wang, R.~Yan, J.~Zhang, S.~Han, G.~Huang, and L.~Zhang, ``Precise condition synthesis for program repair,'' in \emph{2017 IEEE/ACM 39th International Conference on Software Engineering (ICSE)}.\hskip 1em plus 0.5em minus 0.4em\relax IEEE, 2017, pp. 416--426.

\bibitem{long2017automatic}
F.~Long, P.~Amidon, and M.~Rinard, ``Automatic inference of code transforms for patch generation,'' in \emph{Proceedings of the 2017 11th Joint Meeting on Foundations of Software Engineering}, 2017, pp. 727--739.

\bibitem{hua2018towards}
J.~Hua, M.~Zhang, K.~Wang, and S.~Khurshid, ``Towards practical program repair with on-demand candidate generation,'' in \emph{Proceedings of the 40th international conference on software engineering}, 2018, pp. 12--23.

\bibitem{wen2018context}
M.~Wen, J.~Chen, R.~Wu, D.~Hao, and S.-C. Cheung, ``Context-aware patch generation for better automated program repair,'' in \emph{Proceedings of the 40th international conference on software engineering}, 2018, pp. 1--11.

\bibitem{jiang2018shaping}
J.~Jiang, Y.~Xiong, H.~Zhang, Q.~Gao, and X.~Chen, ``Shaping program repair space with existing patches and similar code,'' in \emph{Proceedings of the 27th ACM SIGSOFT international symposium on software testing and analysis}, 2018, pp. 298--309.

\bibitem{liu2019tbar}
K.~Liu, A.~Koyuncu, D.~Kim, and T.~F. Bissyand{\'e}, ``Tbar: Revisiting template-based automated program repair,'' in \emph{Proceedings of the 28th ACM SIGSOFT international symposium on software testing and analysis}, 2019, pp. 31--42.

\bibitem{liu2019avatar}
------, ``Avatar: Fixing semantic bugs with fix patterns of static analysis violations,'' in \emph{2019 IEEE 26th International Conference on Software Analysis, Evolution and Reengineering (SANER)}.\hskip 1em plus 0.5em minus 0.4em\relax IEEE, 2019, pp. 1--12.

\bibitem{ghanbari2019practical}
A.~Ghanbari, S.~Benton, and L.~Zhang, ``Practical program repair via bytecode mutation,'' in \emph{Proceedings of the 28th ACM SIGSOFT International Symposium on Software Testing and Analysis}, 2019, pp. 19--30.

\bibitem{jiang2019inferring}
J.~Jiang, L.~Ren, Y.~Xiong, and L.~Zhang, ``Inferring program transformations from singular examples via big code,'' in \emph{2019 34th IEEE/ACM International Conference on Automated Software Engineering (ASE)}.\hskip 1em plus 0.5em minus 0.4em\relax IEEE, 2019, pp. 255--266.

\bibitem{xia2023plastic}
C.~S. Xia, Y.~Ding, and L.~Zhang, ``The plastic surgery hypothesis in the era of large language models,'' in \emph{2023 38th IEEE/ACM International Conference on Automated Software Engineering (ASE)}.\hskip 1em plus 0.5em minus 0.4em\relax IEEE, 2023, pp. 522--534.

\bibitem{chen2019sequencer}
Z.~Chen, S.~Kommrusch, M.~Tufano, L.-N. Pouchet, D.~Poshyvanyk, and M.~Monperrus, ``Sequencer: Sequence-to-sequence learning for end-to-end program repair,'' \emph{IEEE Transactions on Software Engineering}, vol.~47, no.~9, pp. 1943--1959, 2019.

\bibitem{lutellier2020coconut}
T.~Lutellier, H.~V. Pham, L.~Pang, Y.~Li, M.~Wei, and L.~Tan, ``Coconut: combining context-aware neural translation models using ensemble for program repair,'' in \emph{Proceedings of the 29th ACM SIGSOFT international symposium on software testing and analysis}, 2020, pp. 101--114.

\bibitem{li2020dlfix}
Y.~Li, S.~Wang, and T.~N. Nguyen, ``Dlfix: Context-based code transformation learning for automated program repair,'' in \emph{Proceedings of the ACM/IEEE 42nd international conference on software engineering}, 2020, pp. 602--614.

\bibitem{zhu2021syntax}
Q.~Zhu, Z.~Sun, Y.-a. Xiao, W.~Zhang, K.~Yuan, Y.~Xiong, and L.~Zhang, ``A syntax-guided edit decoder for neural program repair,'' in \emph{Proceedings of the 29th ACM joint meeting on European software engineering conference and symposium on the foundations of software engineering}, 2021, pp. 341--353.

\bibitem{jiang2021cure}
N.~Jiang, T.~Lutellier, and L.~Tan, ``Cure: Code-aware neural machine translation for automatic program repair,'' in \emph{2021 IEEE/ACM 43rd International Conference on Software Engineering (ICSE)}.\hskip 1em plus 0.5em minus 0.4em\relax IEEE, 2021, pp. 1161--1173.

\bibitem{ye2022neural}
H.~Ye, M.~Martinez, and M.~Monperrus, ``Neural program repair with execution-based backpropagation,'' in \emph{Proceedings of the 44th international conference on software engineering}, 2022, pp. 1506--1518.

\bibitem{zhu2023tare}
Q.~Zhu, Z.~Sun, W.~Zhang, Y.~Xiong, and L.~Zhang, ``Tare: Type-aware neural program repair,'' in \emph{2023 IEEE/ACM 45th International Conference on Software Engineering (ICSE)}.\hskip 1em plus 0.5em minus 0.4em\relax IEEE, 2023, pp. 1443--1455.

\bibitem{fu2022vulrepair}
M.~Fu, C.~Tantithamthavorn, T.~Le, V.~Nguyen, and D.~Phung, ``Vulrepair: a t5-based automated software vulnerability repair,'' in \emph{Proceedings of the 30th ACM joint european software engineering conference and symposium on the foundations of software engineering}, 2022, pp. 935--947.

\bibitem{Xia2022LessTM}
\BIBentryALTinterwordspacing
C.~Xia and L.~Zhang, ``Less training, more repairing please: revisiting automated program repair via zero-shot learning,'' \emph{Proceedings of the 30th ACM Joint European Software Engineering Conference and Symposium on the Foundations of Software Engineering}, 2022. [Online]. Available: \url{https://api.semanticscholar.org/CorpusID:250627519}
\BIBentrySTDinterwordspacing

\bibitem{feng2024prompting}
S.~Feng and C.~Chen, ``Prompting is all you need: Automated android bug replay with large language models,'' in \emph{Proceedings of the 46th IEEE/ACM International Conference on Software Engineering}, 2024, pp. 1--13.

\bibitem{guo2022unixcoder}
D.~Guo, S.~Lu, N.~Duan, Y.~Wang, M.~Zhou, and J.~Yin, ``Unixcoder: Unified cross-modal pre-training for code representation,'' \emph{arXiv preprint arXiv:2203.03850}, 2022.

\bibitem{nijkamp2022codegen}
E.~Nijkamp, B.~Pang, H.~Hayashi, L.~Tu, H.~Wang, Y.~Zhou, S.~Savarese, and C.~Xiong, ``Codegen: An open large language model for code with multi-turn program synthesis,'' \emph{arXiv preprint arXiv:2203.13474}, 2022.

\bibitem{wang2023codet5+}
Y.~Wang, H.~Le, A.~D. Gotmare, N.~D. Bui, J.~Li, and S.~C. Hoi, ``Codet5+: Open code large language models for code understanding and generation,'' \emph{arXiv preprint arXiv:2305.07922}, 2023.

\bibitem{ahmed2022few}
T.~Ahmed and P.~Devanbu, ``Few-shot training llms for project-specific code-summarization,'' in \emph{Proceedings of the 37th IEEE/ACM International Conference on Automated Software Engineering}, 2022, pp. 1--5.

\bibitem{prenner2022can}
J.~A. Prenner, H.~Babii, and R.~Robbes, ``Can openai's codex fix bugs? an evaluation on quixbugs,'' in \emph{Proceedings of the Third International Workshop on Automated Program Repair}, 2022, pp. 69--75.

\bibitem{jiang2023impact}
N.~Jiang, K.~Liu, T.~Lutellier, and L.~Tan, ``Impact of code language models on automated program repair,'' in \emph{2023 IEEE/ACM 45th International Conference on Software Engineering (ICSE)}.\hskip 1em plus 0.5em minus 0.4em\relax IEEE, 2023, pp. 1430--1442.

\bibitem{xia2023automated}
C.~S. Xia, Y.~Wei, and L.~Zhang, ``Automated program repair in the era of large pre-trained language models,'' in \emph{Proceedings of the 45th International Conference on Software Engineering (ICSE 2023). Association for Computing Machinery}, 2023.

\bibitem{fan2023automated}
Z.~Fan, X.~Gao, M.~Mirchev, A.~Roychoudhury, and S.~H. Tan, ``Automated repair of programs from large language models,'' in \emph{2023 IEEE/ACM 45th International Conference on Software Engineering (ICSE)}.\hskip 1em plus 0.5em minus 0.4em\relax IEEE, 2023, pp. 1469--1481.

\bibitem{zhang2023gamma}
Q.~Zhang, C.~Fang, T.~Zhang, B.~Yu, W.~Sun, and Z.~Chen, ``Gamma: Revisiting template-based automated program repair via mask prediction,'' in \emph{2023 38th IEEE/ACM International Conference on Automated Software Engineering (ASE)}.\hskip 1em plus 0.5em minus 0.4em\relax IEEE, 2023, pp. 535--547.

\bibitem{yin2024thinkrepair}
X.~Yin, C.~Ni, S.~Wang, Z.~Li, L.~Zeng, and X.~Yang, ``Thinkrepair: Self-directed automated program repair,'' in \emph{Proceedings of the 33rd ACM SIGSOFT International Symposium on Software Testing and Analysis}, 2024, pp. 1274--1286.

\bibitem{xia2023keep}
C.~S. Xia and L.~Zhang, ``Keep the conversation going: Fixing 162 out of 337 bugs for \$0.42 each using chatgpt,'' \emph{arXiv preprint arXiv:2304.00385}, 2023.

\bibitem{huang2023empirical}
K.~Huang, X.~Meng, J.~Zhang, Y.~Liu, W.~Wang, S.~Li, and Y.~Zhang, ``An empirical study on fine-tuning large language models of code for automated program repair,'' in \emph{2023 38th IEEE/ACM International Conference on Automated Software Engineering (ASE)}.\hskip 1em plus 0.5em minus 0.4em\relax IEEE, 2023, pp. 1162--1174.

\bibitem{xiang2024far}
J.~Xiang, X.~Xu, F.~Kong, M.~Wu, Z.~Zhang, H.~Zhang, and Y.~Zhang, ``How far can we go with practical function-level program repair?'' \emph{arXiv preprint arXiv:2404.12833}, 2024.

\bibitem{feng2020codebert}
Z.~Feng, D.~Guo, D.~Tang, N.~Duan, X.~Feng, M.~Gong, L.~Shou, B.~Qin, T.~Liu, D.~Jiang \emph{et~al.}, ``Codebert: A pre-trained model for programming and natural languages,'' \emph{arXiv preprint arXiv:2002.08155}, 2020.

\bibitem{wang2021codet5}
Y.~Wang, W.~Wang, S.~Joty, and S.~C. Hoi, ``Codet5: Identifier-aware unified pre-trained encoder-decoder models for code understanding and generation,'' \emph{arXiv preprint arXiv:2109.00859}, 2021.

\bibitem{just2014defects4j}
R.~Just, D.~Jalali, and M.~D. Ernst, ``Defects4j: A database of existing faults to enable controlled testing studies for java programs,'' in \emph{Proceedings of the 2014 international symposium on software testing and analysis}, 2014, pp. 437--440.

\bibitem{le2015manybugs}
C.~Le~Goues, N.~Holtschulte, E.~K. Smith, Y.~Brun, P.~Devanbu, S.~Forrest, and W.~Weimer, ``The manybugs and introclass benchmarks for automated repair of c programs,'' \emph{IEEE Transactions on Software Engineering}, vol.~41, no.~12, pp. 1236--1256, 2015.

\bibitem{lin2017quixbugs}
D.~Lin, J.~Koppel, A.~Chen, and A.~Solar-Lezama, ``Quixbugs: A multi-lingual program repair benchmark set based on the quixey challenge,'' in \emph{Proceedings Companion of the 2017 ACM SIGPLAN international conference on systems, programming, languages, and applications: software for humanity}, 2017, pp. 55--56.

\bibitem{tu2024overview}
X.~Tu, Z.~He, Y.~Huang, Z.-H. Zhang, M.~Yang, and J.~Zhao, ``An overview of large ai models and their applications,'' \emph{Visual Intelligence}, vol.~2, no.~1, pp. 1--22, 2024.

\bibitem{sharir2020cost}
O.~Sharir, B.~Peleg, and Y.~Shoham, ``The cost of training nlp models: A concise overview,'' \emph{arXiv preprint arXiv:2004.08900}, 2020.

\bibitem{roziere2023code}
B.~Roziere, J.~Gehring, F.~Gloeckle, S.~Sootla, I.~Gat, X.~E. Tan, Y.~Adi, J.~Liu, T.~Remez, J.~Rapin \emph{et~al.}, ``Code llama: Open foundation models for code,'' \emph{arXiv preprint arXiv:2308.12950}, 2023.

\bibitem{touvron2023llama}
H.~Touvron, L.~Martin, K.~Stone, P.~Albert, A.~Almahairi, Y.~Babaei, N.~Bashlykov, S.~Batra, P.~Bhargava, S.~Bhosale \emph{et~al.}, ``Llama 2: Open foundation and fine-tuned chat models,'' \emph{arXiv preprint arXiv:2307.09288}, 2023.

\bibitem{li2023starcoder}
R.~Li, L.~B. Allal, Y.~Zi, N.~Muennighoff, D.~Kocetkov, C.~Mou, M.~Marone, C.~Akiki, J.~Li, J.~Chim \emph{et~al.}, ``Starcoder: may the source be with you!'' \emph{arXiv preprint arXiv:2305.06161}, 2023.

\bibitem{guo2024deepseekcoderlargelanguagemodel}
\BIBentryALTinterwordspacing
D.~Guo, Q.~Zhu, D.~Yang, Z.~Xie, K.~Dong, W.~Zhang, G.~Chen, X.~Bi, Y.~Wu, Y.~K. Li, F.~Luo, Y.~Xiong, and W.~Liang, ``Deepseek-coder: When the large language model meets programming -- the rise of code intelligence,'' 2024. [Online]. Available: \url{https://arxiv.org/abs/2401.14196}
\BIBentrySTDinterwordspacing

\bibitem{10298287}
G.~An, M.~Kwon, K.~Choi, J.~Yi, and S.~Yoo, ``Bugsc++: A highly usable real world defect benchmark for c/c++,'' in \emph{2023 38th IEEE/ACM International Conference on Automated Software Engineering (ASE)}, 2023, pp. 2034--2037.

\bibitem{durieux2016introclassjava}
T.~Durieux and M.~Monperrus, ``Introclassjava: A benchmark of 297 small and buggy java programs,'' Ph.D. dissertation, Universite Lille 1, 2016.

\bibitem{wu2023ConDefectsnewdatasetaddress}
\BIBentryALTinterwordspacing
Y.~Wu, Z.~Li, J.~M. Zhang, and Y.~Liu, ``Condefects: A new dataset to address the data leakage concern for llm-based fault localization and program repair,'' 2023. [Online]. Available: \url{https://arxiv.org/abs/2310.16253}
\BIBentrySTDinterwordspacing

\bibitem{10.5555/3295222.3295349}
A.~Vaswani, N.~Shazeer, N.~Parmar, J.~Uszkoreit, L.~Jones, A.~N. Gomez, L.~Kaiser, and I.~Polosukhin, ``Attention is all you need,'' in \emph{Proceedings of the 31st International Conference on Neural Information Processing Systems}, ser. NIPS'17.\hskip 1em plus 0.5em minus 0.4em\relax Red Hook, NY, USA: Curran Associates Inc., 2017, p. 6000–6010.

\bibitem{kaplan2020scaling}
J.~Kaplan, S.~McCandlish, T.~Henighan, T.~B. Brown, B.~Chess, R.~Child, S.~Gray, A.~Radford, J.~Wu, and D.~Amodei, ``Scaling laws for neural language models,'' \emph{arXiv preprint arXiv:2001.08361}, 2020.

\bibitem{sun2019fine}
C.~Sun, X.~Qiu, Y.~Xu, and X.~Huang, ``How to fine-tune bert for text classification?'' in \emph{China national conference on Chinese computational linguistics}.\hskip 1em plus 0.5em minus 0.4em\relax Springer, 2019, pp. 194--206.

\bibitem{ding2023parameter}
N.~Ding, Y.~Qin, G.~Yang, F.~Wei, Z.~Yang, Y.~Su, S.~Hu, Y.~Chen, C.-M. Chan, W.~Chen \emph{et~al.}, ``Parameter-efficient fine-tuning of large-scale pre-trained language models,'' \emph{Nature Machine Intelligence}, vol.~5, no.~3, pp. 220--235, 2023.

\bibitem{lester2021power}
B.~Lester, R.~Al-Rfou, and N.~Constant, ``The power of scale for parameter-efficient prompt tuning,'' \emph{arXiv preprint arXiv:2104.08691}, 2021.

\bibitem{jia2022visual}
M.~Jia, L.~Tang, B.-C. Chen, C.~Cardie, S.~Belongie, B.~Hariharan, and S.-N. Lim, ``Visual prompt tuning,'' in \emph{European conference on computer vision}.\hskip 1em plus 0.5em minus 0.4em\relax Springer, 2022, pp. 709--727.

\bibitem{silva2023repairllama}
A.~Silva, S.~Fang, and M.~Monperrus, ``Repairllama: Efficient representations and fine-tuned adapters for program repair,'' \emph{arXiv preprint arXiv:2312.15698}, 2023.

\bibitem{hu2022lora}
E.~J. Hu, Y.~Shen, P.~Wallis, Z.~Allen-Zhu, Y.~Li, S.~Wang, L.~Wang, W.~Chen \emph{et~al.}, ``Lora: Low-rank adaptation of large language models.'' \emph{ICLR}, vol.~1, no.~2, p.~3, 2022.

\bibitem{brown2020language}
\BIBentryALTinterwordspacing
T.~B. Brown, B.~Mann, N.~Ryder, M.~Subbiah, J.~Kaplan, P.~Dhariwal, A.~Neelakantan, P.~Shyam, G.~Sastry, A.~Askell, S.~Agarwal, A.~Herbert-Voss, G.~Krueger, T.~Henighan, R.~Child, A.~Ramesh, D.~M. Ziegler, J.~Wu, C.~Winter, C.~Hesse, M.~Chen, E.~Sigler, M.~Litwin, S.~Gray, B.~Chess, J.~Clark, C.~Berner, S.~McCandlish, A.~Radford, I.~Sutskever, and D.~Amodei, ``Language models are few-shot learners,'' in \emph{Advances in Neural Information Processing Systems}, vol.~33, 2020, pp. 1877--1901. [Online]. Available: \url{https://proceedings.neurips.cc/paper/2020/hash/1457c0d6bfcb4967418bfb8ac142f64a-Abstract.html}
\BIBentrySTDinterwordspacing

\bibitem{wei2022chain}
\BIBentryALTinterwordspacing
J.~Wei, X.~Wang, D.~Schuurmans, M.~Bosma, B.~Ichter, F.~Xia, E.~Chi, Q.~V. Le, and D.~Zhou, ``Chain-of-thought prompting elicits reasoning in large language models,'' in \emph{Advances in Neural Information Processing Systems}, vol.~35, 2022. [Online]. Available: \url{https://proceedings.neurips.cc/paper/2022/hash/9d5609613524ecf4f15af0f7b31abca4-Abstract-Conference.html}
\BIBentrySTDinterwordspacing

\bibitem{zhang2022auto}
\BIBentryALTinterwordspacing
Z.~Zhang, A.~Zhang, M.~Li, and A.~Smola, ``Automatic chain of thought prompting in large language models,'' 2022. [Online]. Available: \url{https://arxiv.org/abs/2210.03493}
\BIBentrySTDinterwordspacing

\bibitem{wang2023towards}
\BIBentryALTinterwordspacing
B.~Wang, S.~Min, X.~Deng, J.~Shen, Y.~Wu, L.~Zettlemoyer, and H.~Sun, ``Towards understanding chain-of-thought prompting: An empirical study of what matters,'' in \emph{Proceedings of the 61st Annual Meeting of the Association for Computational Linguistics (Volume 1: Long Papers)}.\hskip 1em plus 0.5em minus 0.4em\relax Toronto, Canada: Association for Computational Linguistics, 2023, pp. 2717--2739. [Online]. Available: \url{https://aclanthology.org/2023.acl-long.153/}
\BIBentrySTDinterwordspacing

\bibitem{wei2023copiloting}
Y.~Wei, C.~S. Xia, and L.~Zhang, ``Copiloting the copilots: Fusing large language models with completion engines for automated program repair,'' in \emph{Proceedings of the 31st ACM Joint European Software Engineering Conference and Symposium on the Foundations of Software Engineering}, 2023, pp. 172--184.

\bibitem{NEURIPS2020_1457c0d6}
\BIBentryALTinterwordspacing
T.~Brown, B.~Mann, N.~Ryder, M.~Subbiah, J.~D. Kaplan, P.~Dhariwal, A.~Neelakantan, P.~Shyam, G.~Sastry, A.~Askell, S.~Agarwal, A.~Herbert-Voss, G.~Krueger, T.~Henighan, R.~Child, A.~Ramesh, D.~Ziegler, J.~Wu, C.~Winter, C.~Hesse, M.~Chen, E.~Sigler, M.~Litwin, S.~Gray, B.~Chess, J.~Clark, C.~Berner, S.~McCandlish, A.~Radford, I.~Sutskever, and D.~Amodei, ``Language models are few-shot learners,'' in \emph{Advances in Neural Information Processing Systems}, H.~Larochelle, M.~Ranzato, R.~Hadsell, M.~Balcan, and H.~Lin, Eds., vol.~33.\hskip 1em plus 0.5em minus 0.4em\relax Curran Associates, Inc., 2020, pp. 1877--1901. [Online]. Available: \url{https://proceedings.neurips.cc/paper_files/paper/2020/file/1457c0d6bfcb4967418bfb8ac142f64a-Paper.pdf}
\BIBentrySTDinterwordspacing

\bibitem{li2025bybrid}
F.~Li, J.~Jiang, J.~Sun, and H.~Zhang, ``Hybrid automated program repair by combining large language models and program analysis,'' \emph{ACM Trans. Softw. Eng. Methodol.}, Jan. 2025, just Accepted.

\bibitem{atcoder}
``Atcoder,'' \url{https://atcoder.jp}.

\bibitem{ouyang2024benchmarking}
Y.~Ouyang, J.~Yang, and L.~Zhang, ``Benchmarking automated program repair: An extensive study on both real-world and artificial bugs,'' in \emph{Proceedings of the 33rd ACM SIGSOFT International Symposium on Software Testing and Analysis}, 2024, pp. 440--452.

\bibitem{yang2023large}
J.~Yang, Y.~Wang, Y.~Lou, M.~Wen, and L.~Zhang, ``A large-scale empirical review of patch correctness checking approaches,'' in \emph{Proceedings of the 31st ACM Joint European Software Engineering Conference and Symposium on the Foundations of Software Engineering}, 2023, pp. 1203--1215.

\bibitem{falleri2014fine}
J.-R. Falleri, F.~Morandat, X.~Blanc, M.~Martinez, and M.~Monperrus, ``Fine-grained and accurate source code differencing,'' in \emph{Proceedings of the 29th ACM/IEEE international conference on Automated software engineering}, 2014, pp. 313--324.

\bibitem{martinez2023hyperparameter}
M.~Martinez, J.-R. Falleri, and M.~Monperrus, ``Hyperparameter optimization for ast differencing,'' \emph{IEEE Transactions on Software Engineering}, vol.~49, no.~10, pp. 4814--4828, 2023.

\end{thebibliography}
